\documentclass[11pt]{article} 

\usepackage{amsmath,amsthm,latexsym,amssymb,amsfonts,epsfig}

\usepackage{cite}
\usepackage{hyperref}
\hypersetup{
    colorlinks,%
    citecolor=blue,%
    filecolor=blue,%
    linkcolor=blue,%
    urlcolor=blue
}

\newcommand{\be}{\begin{equation}}
\newcommand{\ee}{\end{equation}}
\newcommand{\ba}{\begin{eqnarray}}
\newcommand{\ea}{\end{eqnarray}}

\title{{\sf Hamiltonian Renormalisation II.}\\
{\sf Renormalisation Flow of 1+1 dimensional free scalar fields:}\\
{\sf Derivation}
} 
\author{
{\sf T. Lang}$^1$\thanks{{\sf 
thorsten.lang@gravity.fau.de}},
{\sf K. Liegener}$^1$\thanks{{\sf 
klaus.liegener@gravity.fau.de}},
{\sf T. Thiemann}$^1$\thanks{{\sf 
thomas.thiemann@gravity.fau.de}}\\
\\
{\sf $^1$ Inst. for Quantum Gravity, FAU Erlangen -- N\"urnberg,}\\
{\sf Staudtstr. 7, 91058 Erlangen, Germany}\\
}
\date{{\small\sf \today}}

\makeatletter
\@addtoreset{equation}{section}
\makeatother

\begin{document} 

\maketitle

{\sf
\begin{abstract}
In the companion paper \cite{1} we motivated a renormalisation flow
on Osterwalder-Schrader data (OS-data) consisting of 1. a Hilbert space,
2. a cyclic vacuum and 3. a Hamiltonian annihilating that vacuum.
As the name suggests, the motivation was via the OS reconstruction
theorem which allows to reconstruct the OS data from an OS measure satisfying
(a subset of) the OS axioms, in particular reflection positivity. 
The guiding principle was to map the usual Wilsonian 
path integral renormalisation flow onto a flow of the corresponding OS 
data. 

We showed that this induced flow on the OS data has an unwanted 
feature which disqualifies the associated coarse grained Hamiltonians
from being the projections of a continuum Hamiltonian onto vectors in the 
coarse grained Hilbert space. This motivated the definition of a direct
Hamiltonian renormalisation flow which follows the guiding principle 
but does not suffer from the afore mentioned caveat.

In order to test our proposal, we apply it to the only known completely solvable 
model, namely the case of free scalar quantum fields. In this paper we focus
on the Klein Gordon field in two spacetime dimensions and illustrate the
difference between the path integral induced and direct Hamiltonian 
flow. Generalisations to more general models in higher dimensions will
be discussed in our companion papers. 
\end{abstract}

\newpage

\tableofcontents

\newpage

~\\
{\bf Notation:}\\
\\
In this paper we will deal with quantum fields in the presence of an 
infrared cut-off $R$ and with smearing functions of finite time support 
in $[-T,T]$. The spatial ultraviolet cut-off is denoted by $M$ and has the  interpretation of the number of lattice vertices in each spatial 
direction. We will mostly not be interested in an analogous temporal 
ultraviolet cut-off $N$ but sometimes refer to it for comparison 
with other approaches. These quantities allow us to define dimensionful
cut-offs $\epsilon_{RM}=\frac{R}{M},\; \delta_{TN}=\frac{T}{N}$. In 
Fourier space we define analogously 
$k_R=\frac{2\pi}{R},\; k_M=\frac{2\pi}{M},\;   
k_T=\frac{2\pi}{T},\; k_N=\frac{2\pi}{N}$.

We will deal with both instantaneous fields, smearing functions and Weyl 
elements as well as corresponding temporally dependent objects. 
The instantaneous objects are denoted by lower case letters 
$\phi_{RM}, \; f_{RM},\; w_{RM}[f_{RM}]$, the temporally dependent ones by
upper case ones
$\Phi_{RM}, \; F_{RM},\; W_{RM}[F_{RM}]$. As we will see, smearing functions
$F_{RM}$ with 
compact and discrete (sharp) time support will play a more fundamental role 
for our purposes than those with a smoother dependence. 

Osterwalder-Schrader reconstruction concerns the interplay between 
time translation invariant, time reflection invariant and reflection positive 
measures (OS measures)
$\mu_{RM}$ on the space of history fields $\Phi_{RM}$ and their corresponding
Osterwalder-Schrader (OS) data ${\cal H}_{RM}, \Omega_{RM}, H_{RM}$ where   
${\cal H}_{RM}$ is a Hilbert space with cyclic (vacuum) vector $\Omega_{RM}$ 
annihilated by a self-adjoint Hamiltonian $H_{RM}$. Together, the vector
$\Omega_{RM}$ and the scalar product $<.,.>_{{\cal H}_{RM}}$ define 
a measure $\nu_{RM}$ on the space of instantaneous fields $\phi_{RM}$. 

Renormalisation consists in defining a flow or sequence 
$n\to \mu^{(n)}_{RM},\; n\in \mathbb{N}_0$ for all
$M$ of families of measures $\{\mu^{(n)}_{RM}\}_{M\in \mathbb{N}}$. 
The flow will be defined in terms of a coarse graining or embedding 
map $I_{RM\to M'},\; M<M'$ acting on the smearing functions and satisfying 
certain properties that will grant that 1. the resulting fixed point family 
of measures, if it exists, is cylindrically consistent and 2. the flow 
stays within the class of OS measures. Fixed point quantities are denoted 
by an upper case $^\ast$, e.g. $\mu^\ast_{RM}$.

\newpage

\section{Introduction}
\label{s1}

In our companion paper \cite{1} we motivated a renormalisation flow 
directly on the Osterwalder-Schrader data (OS data) 
consisting of a Hilbert space $\cal H$ supporting a 
Hamiltonian $H$ that annihilates a cyclic vacuum vector $\Omega$ therein.
The motivation was through the usual Wilsonian renormalisation 
flow on OS measures $\mu$ satisfying at least a subset of the OS axioms   
\cite{OS72, GJ87}, specifically time translation invariance, time reflection 
invariance and time reflection positivity. The OS reconstruction 
theorem allows to derive OS data directly from an associated OS measure 
$\mu$ and thus allows to translate the Wilsonian renormalisation flow 
on $\mu$ directly in terms of the OS data. 

The reason for the interest in the flow of the OS data rather than 
of the OS measure stems from applications where the construction of 
the interacting quantum field theory (QFT) \cite{Fr78,Riv00,Sim74} has a more 
natural starting point in its 
Hamiltonian formulation and where the construction of the corresponding 
OS measure is much more difficult. This applies in particular to 
realistic, i.e. Lorentzian signature, quantum
gravity in its canonical formulation \cite{Ash91,Rov04,Thi07,GS13} in the presence of 
matter \cite{KT91,BK95,HP11,GT12} which allows to gauge fix the spacetime diffeomorphism 
invariance and to define a physical Hamiltonian. Due to the complex, not even 
polynomial, interaction in this gravitational Hamiltonian and the fact 
that it does not have a natural (that is, without breaking 
background independence of Einstein's theory) split into a free and an 
interacting piece, the  
path integral measure corresponding to it cannot be simply constructed using 
the usual techniques based on Gaussian measures \cite{GJ87} for QFT on 
Minkowski space and thus presents
a major technical challenge. In other words, if one artificially introduces 
such a split using a background metric, one must make sure that the 
final result of the Wilsonian flow \cite{WK73,Wil75} is independent of that background.
It is widely believed that the Gaussian fixed point of the associated 
flow does not define a renormalisable theory, however, there maybe 
non-Gaussian fixed points that do.  
See \cite{ReuSau07,Perc10,ReuSau12,Eich17} for the state of the art of this so-called ``Asymptotic 
Safety Approach'' to quantum gravity which, to the best knowledge of the 
present authors, is best understood in the Euclidian signature regime.

The renormalisation flow of the path integral measure 
is defined in terms of a coarse graining 
map (or block spin transformation in statistical physics jargon) acting 
on the smearing fields of the spacetime quantum fields in question.
This is because the measure can be defined in terms of its generating 
functional (i.e. the Fourier transform of the measure) which employs 
(generalised) exponentials of the smeared fields. The smearing 
fields carry a label $M$ that specifies its spatial resolution. We restrict 
to spatial resolution and do not consider temporal resolution because 
we want to monitor the induced flow of the OS data which requires that time 
is continuous in order that OS reconstruction applies. 
When restricting to integrate quantum fields 
smeared with fields of given resolution $M$, one obtains 
a whole family of measures $\mu_M$. The coarse graining 
map $I_{M\to M'}$ embeds the space of smearing functions with coarse spatial 
resolution 
$M$ into a space of smearing function with finer spatial resolution $M'>M$.
The flow then constructs a sequence $n\mapsto \{\mu^{(n)}_{M}\}$ of measure
families where $\mu^{(n+1)}_{M'}$ is defined in terms of $\mu^{(n)}_{M},\;
M'>M$. We showed that this sequence stays of OS type if the initial 
family is. By OS reconstruction we then obtain a corresponding 
OS data family $({\cal H}^{(n)}_{M},\;\Omega^{(n)}_{M}, \; H^{(n)}_{M})$.

In \cite{1} we demonstrated that the path integral induced flow of the 
OS data has a peculiar feature. Namely, the $H^{(n)}_{M'}$ commute with the 
projection maps $P^{(n)}_{M\to M'}:\;{\cal H}^{(n)}_{M'}\to {\cal H}^{(n)}_{M'},\;
M<M'$ onto the image of ${\cal H}^{(n+1)}_M$ in ${\cal H}^{(n)}_{M'}$. 
This means that the corresponding fixed point objects have the property 
$[H^\ast_{M'},P^\ast_{M\to M'}]=0$. Hence, the fine resolution 
Hamiltonian preserves every coarse resolution subspace. This is counter
intuitive and contradicts properties of such Hamiltonians in concrete 
applications, e.g. in lattice gauge theory where $M$ defines the lattice 
spacing and where the Hamiltonian will have non-vanishing matrix elements 
between states defined in terms of coarse and fine resolution plaquettes 
respectively. 

In \cite{1} we found that the underlying reason for this peculiarity is that 
the reconstructed OS data are automatically consistent with the semigroup 
law of the OS contraction semi-group. This means that the Hamiltonians 
$H^\ast_{M}$ cannot be interpreted as restrictions of the continuum Hamiltonian 
$H^\ast=\lim_{M\to \infty} H^\ast_{M}$ to ${\cal H}^\ast_{M}$. Thus, the path 
integral induced flow of the OS data is not what we wished to have. This 
motivated us to minimally modify the path integral induced flow  and define a 
direct Hamiltonian flow of OS data which does not need the path integral at 
all. This flow can be considered 1. as a flow of the canonical Hilbert space 
measures $\nu^{(n)}_M$ underlying ${\cal H}^{(n)}_M$ which carries an imprint
of the $\mu^{(n)}_M$ in the sense that $\nu^{(n)}_M$ is the restriction 
of $\mu^{(n)}_M$ to sharp time zero smearing fields and 2. of the Hamiltonians
$H^{(n)}_M$ which keep track of the flow of vacua $\Omega^{(n)}_M$. 
The fixed point of that flow, if it exists and if certain mild 
universality assumptions hold, is an inductive limit of Hilbert   
spaces ${\cal H}^\ast$ together with a densely defined, positive quadratic 
form thereon which in fortunate cases has a Friedrichs extension as a 
positive self-adjoint operator. The direct and path integral induced flow will 
thus be very different in general and the interesting question is whether the 
continuum theories at the fixed points agree with each other. 

In this paper we would like to see the formalism in action and apply it 
to the two-dimensional Klein-Gordon field after having specialised the general
theory to free scalar fields in any dimension. The restriction to two 
dimensions is mainly for illustrative reasons and in order to have a 
concrete example in mind. The free field case serves as a testing ground for
our formalism and allows us to solve all equations analytically.
We find explicitly both the path integral induced and direct Hamiltonian
flow and can determine their fixed point families. Their flows and fixed 
point families are very different from each other. Yet, their continuum 
theories agree. The way this happens is very interesting. At any 
finite resolution and at the fixed point the path integral induced 
Hamiltonian is described by an infinite tower of sharp time zero fields species
and their conjugate momenta with resolution dependent masses. The Hamiltonian 
is quadratic in these fields but they are mutually interacting with resolution
depending coupling ``constants''. By contrast, the direct Hamiltonian 
at any finite resolution and at the fixed point is described by a 
single field species and its conjugate momentum with resolution dependent
mass term. Now, in the continuum limit $M\to \infty$ for the path integral 
induced Hamiltonian all but a single coupling
goes to zero and all but a single corresponding mass goes to infinity.  
This continuum limit agrees with that of the direct Hamiltonian. Thus,
the direct renormalisation flow {\it does} find the correct continuum theory,
at least in the free field case,
and the finite resolution Hamiltonians are the projections to finite 
resolution subspaces of the continuum Hilbert space, which is what we 
wanted to achieve. The direct scheme no longer needs any path integral 
techniques which is what we were after. 
\\
The architecture of this paper is as follows:\\
\\

In section \ref{s5} we review the continuum formulation of a free scalar
field starting from the classical action. That is, we derive its classical
Hamiltonian formulation, the Fock space quantisation, the OS construction --
that is, the derivation of 
the corresponding Wiener measure (Mehler kernel) -- and finally the OS 
reconstruction of the Fock space OS data starting from that Wiener measure.     

In section \ref{s7} we introduce an initial family of OS data labelled 
by a finite resolution parameter based on a certain naive discretisation 
of the continuum theory and its corresponding Fock quantisation. 
This serves to construct an initial family of Wiener measures. We then 
compute both the path integral induced flow and the direct flow of these
OS data for the concrete model of the 1+1-dimensional Klein Gordon field 
and compare between them.

In section \ref{s8} we summarise our findings.\\
\\
We will continue the investigation of the present model in our companion paper
\cite{6} with respect to the aspects of universality of the fixed 
point (i.e. its dependence on the initial values of the OS data and 
of the coarse graining map), its convergence behaviour to 
the fixed point and the decay properties of the contributions 
of the n-th order neighbour points of the improved and perfect 
lattice Laplacian.

The restriction to two dimensions forbids the analysis of the rotational 
invariance of the renormalisation flow, i.e. the question whether 
it is possible to show at finite resolution that the flow converges to
a rotationally invariant fixed point. This will be remedied in our
companion paper \cite{7} where we show that the findings of this paper and 
\cite{6} easily generalise to higher dimensions. It turns out that it is 
sufficient to investigate this for the 2+1-dimensional case due to 
certain factorisation properties. The analysis of rotational invariance
is a showcase for the important question how renormalisation restores 
symmetries of the classical continuum theory when it is broken by the 
initial choice of discretisations, a question which is especially 
important in quantum gravity \cite{BD09,BDS11,DMS14,BS17}.

\section{Review: OS (re)construction of the 
free quantum scalar field in the continuum}
\label{s5}

In this section we review OS reconstruction and OS construction 
for the free quantum scalar field in the continuum, that is, on Minkowski
space without IR cut-off $R$.
Experts on the subject can safely skip this section.

\subsection{Theory class}
\label{s5.0}
 
We  consider the following class of actions
\be \label{5.1}
S:=\frac{1}{2\kappa}\int_{\mathbb{R}^{D+1}}\;dt\; 
d^Dx\;[\frac{1}{c}\dot{\phi}^2-c\phi\omega^2\phi]
\ee
where $\omega^2$ is any function of $p,\Delta$ which has dimension of inverse 
length, for instance
\be \label{5.1a}
\omega^2=\frac{1}{p^{2(n-1)}}
(-\Delta+p^2)^n,\;n=1,2,...
\ee
which is not Poincar\'e invariant unless $n=1$. 
Here $x^0=ct$ has dimension of length
and $\hbar\kappa$ has dimension cm$^{D-1}$ if $\phi$ is dimension free.
Furthermore, $\Delta=\sum_{a=1}^D (\partial/\partial x^a)^2$ is the Laplacian 
and 
$p$ is the inverse of the Compton wave length. 
The Legendre transform of (\ref{5.1}) results in the Hamiltonian
\be \label{5.1b}
H:=\frac{1}{2}\int_{\mathbb{R}^D}\; d^Dx\;[\kappa c\pi^2+\frac{c}{\kappa}
\phi\omega^2\phi]
\ee
where $\pi=\dot{\phi}/(\kappa c)$ is the momentum conjugate to $\phi$
which has the dimension of an action divided by length$^D$. We will see that 
our methods work for any choice of $\omega^2$ and in any dimension. 
Below, for concreteness we consider the simplest case $D=1$ and 
the Poincar\'e invariant choice $n=1$ in 
(\ref{5.1a}). The general choices for $D,n$ will be considered in our 
companion papers \cite{6,7}.

\subsection{Canonical quantisation}
\label{s5.1}

We choose a 
Fock representation $\cal H$
based on the annihilation operator valued distribution
\be \label{5.3}
a=\frac{1}{\sqrt{2\hbar\kappa}}[\sqrt{\omega}\phi
-i\kappa\frac{1}{\sqrt{\omega}}\pi]
\ee
in terms of which the normal ordered Hamiltonian reads
\be \label{5.4}
H=\hbar c\int\; d^Dx\; a^\ast\; \omega\; a
\ee
The one particle Hilbert space is $L=L_2(\mathbb{R}^D,d^Dx)$. If we denote 
the Fock vacuum by $\Omega$ then  
\be \label{5.4a}
a[f]\Omega=0=H\Omega,\; a[f]:=<f,a>_L=\int_{\mathbb{R}^D}\;d^Dx\; \bar{f}\;a
\ee
and we verify the canonical commutation relations 
\be \label{5.5}
{[}a[f],(a[f'])^\ast]=<f,f'>_L \; 1_{{\cal H}}
\ee
The Weyl elements read for real valued smearing 
functions $f,g\in {\cal S}(\mathbb{R}^D)$ which are smooth and of 
rapid decrease (with $f$ of inverse length$^D$ dimension)
\be \label{5.6}
w[f,g]=e^{i(<f,\phi>_L+<g,\pi>_L)},\;w[f]:=w[f,g=0]
\ee
We compute the generating functional of the Hilbert space measure $\nu$
\be \label{5.7}
\nu(w[f])
=<\Omega,w[f]\Omega>_{{\cal H}}  
=<\Omega,e^{i\sqrt{\frac{\hbar\kappa }{2}}(a[\omega^{-1/2} f]^\ast
+a[\omega^{-1/2} f])}\Omega>_{{\cal H}}  
=e^{-\frac{\hbar\kappa}{4} <f,\omega^{-1} f>}
\ee
which displays $\nu$ as a Gaussian measure with support on the tempered 
distributions $\gamma={\cal S}'(\mathbb{R}^D)$ with zero mean and 
covariance
\be \label{5.8}
C=\frac{\hbar\kappa}{2} \omega^{-1}
\ee
Accordingly, ${\cal H}\cong L_2(\gamma,d\nu)$. 

\subsection{Wiener measure from Hamiltonian Formulation}
\label{s5.2}

We compute the corresponding Wiener measure $\mu$ and anticipate its 
support $\Gamma={\cal S}'(\mathbb{R}^{D+1})$. Let 
\be \label{5.9}
W[F]:=e^{i<F,\Phi>},\;\;F\in {\cal S}(\mathbb{R}^{D+1}),\;\;\Phi\in\Gamma,\;\;
<F,\Phi>=\int_{\mathbb{R}^{D+1}}\; d\beta \; d^{D}x\; \bar{F}\; \Phi
\ee
then for $F$ real valued and in the limit of sharp time support
\be \label{5.10}
F(\beta,x)=\sum_{k=1}^N\;\delta(\beta,\beta_k) f_k(x),\;\beta_k<\beta_{k+1},\;
f_k\in {\cal S}(\mathbb{R})
\ee
we have
\be \label{5.11a}
\mu(W[F]):=<\Omega,w[f_N]\; 
e^{-(\beta_N-\beta_{N-1})H/\hbar}\;...\;
e^{-(\beta_2-\beta_1)H/\hbar}\;w[f_1]\;\Omega>_{{\cal H}}
\ee
To compute (\ref{5.11a}) explicitly we determine the analytically
extended Heisenberg field 
\be \label{5.12a}
z(f,\beta):=e^{-\beta H/\hbar} <f,\phi> e^{\beta H/\hbar}
=<{\rm ch}(c\omega\beta)\cdot f,\phi>
-i<\frac{\kappa}{\omega}{\rm sh}(c\omega\beta) \cdot f,\pi>
\ee
which can be found by analytic continuation $t\to -i\beta$
of its unitary evolution with respect to $H$.
It follows that 
\be \label{5.13a}
e^{-\beta H/\hbar} w[f] e^{\beta H/\hbar}=e^{i\;z(f,\beta)}
\ee
so that, using the vacuum property and the abbreviation 
$K_\beta=e^{-\beta H/\hbar}$ 
\ba \label{5.11}
\mu(W[F]) &=
&<\Omega,w[f_N]\; 
K_{\beta_N-\beta_{N-1}}\;w[f_{N-1}]\;...\;w[f_2]
K_{\beta_2-\beta_1}\;w[f_1]\;\Omega>_{{\cal H}}
\nonumber\\
&=&
<\Omega,w[f_N]\; 
(K_{\beta_N-\beta_{N-1}}\;w[f_{N-1}]\;K_{\beta_N-\beta_{N-1}}^{-1})\;
(K_{\beta_N-\beta_{N-2}}\;w[f_{N-2}]\;K_{\beta_N-\beta_{N-2}}^{-1})\;
...\;
\nonumber\\
&&...
(K_{\beta_N-\beta_1}\;w[f_1]\;K_{\beta_N-\beta_1}^{-1})\;
\;\Omega>_{{\cal H}}
\nonumber\\
&=&
<\Omega,
e^{iz(f_N,\beta_N-\beta_N)}\;
e^{iz(f_{N-1},\beta_N-\beta_{N-1})}\;
...\;
e^{iz(f_1,\beta_N-\beta_1)}\;\Omega>_{{\cal H}}
\ea
Let $z_k=z(f_k,\beta_N-\beta_k)$ then by the BCH formula
\ba \label{5.12}
 e^{i z_N}\; ... e^{i z_2}\; e^{i z_1}
&=& e^{i z_N}\; ... e^{i z_3}\; e^{i [z_1+z_2]}\;\; e^{-\frac{1}{2}[z_2,z_1]}
\nonumber\\
&=& e^{i z_N}\; ... e^{i z_4}\; e^{i [z_1+z_2+z_3]}\;\; 
e^{-\frac{1}{2}([z_2,z_1]+[z_3,z_1+z_2])}
\nonumber\\
&=& e^{i\sum_{k=1}^N\; z_k}\;\;
e^{-\frac{1}{2}\sum_{k=2}^{N}\;\sum_{l=1}^{k-1}\; [z_k,z_l]}
\ea
and by the CCR with $\bar{\beta}_k=\beta_N-\beta_k$
\be \label{5.13}
{[}z_k,z_l]=-\hbar 
<{\rm ch}(c\omega\bar{\beta}_k)\cdot f_k,\;
\frac{\kappa}{\omega}{\rm sh}(c\omega\bar{\beta}_l) \cdot f_l>
+\hbar<{\rm ch}(c\omega\bar{\beta}_l)\cdot f_l,\;
\frac{\kappa}{\omega}{\rm sh}(c\omega\bar{\beta}_k) \cdot f_k>
\ee
Next, with the combinations that appear in the single sum exponent of 
(\ref{5.12})
\be \label{5.14}
f:=\sum_{k=1}^N {\rm ch}(c\omega\bar{\beta}_k)\cdot f_k,\;\;
g:=\sum_{k=1}^N \frac{\kappa}{\omega}{\rm sh}(c\omega\bar{\beta}_k)
\cdot f_k
\ee
we decompose into annihilation and creation operators and apply once more
the BCH formula
\ba \label{5.15}
e^{i\phi[f]+i\pi[g]} &=& 
e^{i<\sqrt{\frac{\hbar\kappa}{2\omega}}\cdot f,a+a^\ast>
+i<\sqrt{\frac{\hbar\omega}{2\kappa}}\cdot g,a-a^\ast>}
\nonumber\\
&=&e^{i[<\sqrt{\frac{\hbar\kappa}{2\omega}}\cdot f
+\sqrt{\frac{\hbar\omega}{2\kappa}}\cdot g,a>+
<\sqrt{\frac{\hbar\kappa}{2\omega}}\cdot f
-\sqrt{\frac{\hbar\omega}{2\kappa}}\cdot g,a^\ast>]}
\nonumber\\
&=& e^{i<\sqrt{\frac{\hbar\kappa}{2\omega}}\cdot f
-\sqrt{\frac{\hbar\omega}{2\kappa}}\cdot g,a^\ast>}\;
e^{i<\sqrt{\frac{\hbar\kappa}{2\omega}}\cdot f
+\sqrt{\frac{\hbar\omega}{2\kappa}}\cdot g,a>}\;\;
e^{\frac{1}{2}
[<\sqrt{\frac{\hbar\kappa}{2\omega}}\cdot f
-\sqrt{\frac{\hbar\omega}{2\kappa}}\cdot g,a^\ast>,
<\sqrt{\frac{\hbar\kappa}{2\omega}}\cdot f
+\sqrt{\frac{\hbar\omega}{2\kappa}}\cdot g,a>]}
\nonumber\\
&=& e^{i<\sqrt{\frac{\hbar\kappa}{2\omega}}\cdot f
-\sqrt{\frac{\hbar\omega}{2\kappa}}\cdot g,a^\ast>}\;
e^{i<\sqrt{\frac{\hbar\kappa}{2\omega}}\cdot f
+\sqrt{\frac{\hbar\omega}{2\kappa}}\cdot g,a>}\;\;
e^{-\frac{1}{2}
<\sqrt{\frac{\hbar\kappa}{2\omega}}\cdot f
-\sqrt{\frac{\hbar\omega}{2\kappa}}\cdot g,
\sqrt{\frac{\hbar\kappa}{2\omega}}\cdot f
+\sqrt{\frac{\hbar\omega}{2\kappa}}\cdot g>}
\ea
The last exponent is explicitly 
\ba \label{5.16}
&&-\frac{\hbar\kappa}{4}\sum_{k,l=1}^N\;
<[{\rm ch}(c\omega\bar{\beta}_k)-{\rm sh}(c\omega\bar{\beta}_k)]\omega^{-1/2}
\cdot f_k,
[{\rm ch}(c\omega\bar{\beta}_l)+{\rm sh}(c\omega\bar{\beta}_k)]\omega^{-1/2}
\cdot f_l>
\nonumber\\
&=&-\frac{\hbar\kappa}{4}\sum_{k,l=1}^N\;
<e^{-c\omega\bar{\beta}_k}\cdot f_k,
\omega^{-1}\; e^{c\omega\bar{\beta}_l}\cdot 
f_l> 
\ea
Now we remember (\ref{5.10}) and thus may write (\ref{5.16}) as 
\ba \label{5.17}
&&-\frac{\hbar\kappa}{4}
\int_{\beta_1}^{\beta_N}\; ds\;\int_{\beta_1}^{\beta_N}\; dt
<e^{-c\omega(\beta_N-s)}\cdot F(s),
\omega^{-1}\; e^{c\omega(\beta_N-t)}\cdot 
F(t)> 
\nonumber\\
&=&-\frac{\hbar\kappa}{4}
\int_{\beta_1}^{\beta_N}\; ds\;\int_{\beta_1}^{\beta_N}\; dt
<e^{c\omega(s-t)}\cdot F(s),
\omega^{-1}\; \cdot 
F(t)> 
\ea
where in the last step we used that $e^{-c(\beta_N-s)\omega}$ is a
self-adjoint smoothening operator. Likewise, we can write 
the double sum exponent in (\ref{5.12}) as, using (\ref{5.13}) 
\ba \label{5.18}
&&
-\frac{\hbar}{2}
\int_{\beta_1}^{\beta_N}\; ds\;\int_{\beta_1}^s\; dt
(- 
<{\rm ch}(c\omega(\beta_N-s))\cdot F(s),\;
\frac{\kappa}{\omega}{\rm sh}(c\omega(\beta_N-t)) \cdot F(t)>
\nonumber\\
&& +<{\rm ch}(c\omega(\beta_N-t))\cdot F(t),\;
\frac{\kappa}{\omega}{\rm sh}(c\omega(\beta_N-s)) \cdot F(s)>)
\nonumber\\
&=&
-\frac{\hbar\kappa}{2}
\int_{\beta_1}^{\beta_N}\; ds\;\int_{\beta_1}^s\; dt
(- 
<{\rm sh}(c\omega(\beta_N-t))\;{\rm ch}(c\omega(\beta_N-s))\cdot F(s),\;
 \omega^{-1}\cdot F(t)>
\nonumber\\
&& +<\omega^{-1}\cdot F(t),\;
{\rm ch}(c\omega(\beta_N-t))\;{\rm sh}(c\omega(\beta_N-s)) \cdot F(s)>)
\nonumber\\
&=&
\frac{\hbar\kappa}{2}
\int_{\beta_1}^{\beta_N}\; ds\;\int_{\beta_1}^s\; dt
<[{\rm sh}(c\omega(\beta_N-t))\;{\rm ch}(c\omega(\beta_N-s))
\nonumber\\
&&
-{\rm ch}(c\omega(\beta_N-t))\;{\rm sh}(c\omega(\beta_N-s))] \cdot F(s),\; 
\omega^{-1}\cdot F(t)>
\nonumber\\
&=&
\frac{\hbar\kappa}{2}
\int_{\beta_1}^{\beta_N}\; ds\;\int_{\beta_1}^s\; dt
<{\rm sh}(c\omega[(\beta_N-t)-(\beta_N-s)])
\cdot F(s),\; 
\omega^{-1}\cdot F(t)>
\nonumber\\
&=&
\frac{\hbar\kappa}{2}
\int_{\beta_1}^{\beta_N}\; ds\;\int_{\beta_1}^s\; dt
<{\rm sh}(c\omega(s-t))\cdot F(s),\; 
\omega^{-1}\cdot F(t)>
\ea
where in the second step we used that all operators involved are self-adjoint
and act on their invariant domain and in the second we used the reality of all
functions involved.

We can now put (\ref{5.12}), (\ref{5.17}) and (\ref{5.18}) together
and find, dropping the constraints $\beta_1\le s,t \le \beta_N$ as $F$ 
has this compact time support anyway
\ba \label{5.19}
\mu(W[F]) 
&=&
\exp(-\frac{\hbar\kappa}{4}
[\int\; ds \;\int\;  dt <e^{c\omega(s-t)}\cdot F(s),\omega^{-1} F(t)>
\nonumber\\
&&
-2\int\; ds \;\int_{t\le s}\;  dt\;
<{\rm sh}(c\omega(s-t))\cdot F(s),\; \omega^{-1}\cdot F(t)>])
\nonumber\\
&=&
\exp(-\frac{\hbar\kappa}{4}
[\int\; ds \;\int_{t\ge s}\;  dt <e^{c\omega(s-t)}\cdot F(s),\omega^{-1} F(t)>
\nonumber\\
&& +\int\; ds \;\int_{t\le s}\;  dt\;
(<e^{-c\omega(s-t)}\cdot F(s),\omega^{-1} F(t)>
\nonumber\\
&&
-<e^{c\omega(s-t)}\cdot F(s),\; \omega^{-1}\cdot F(t)>
\nonumber\\
&& +<e^{c\omega(s-t)}\cdot F(s),\; \omega^{-1}\cdot F(t)>]	)
\nonumber\\
&=&
\exp(-\frac{\hbar\kappa}{4}
[\int\; ds \;\int_{t\ge s}\;  dt <e^{-c\omega(t-s)}\cdot F(s),\omega^{-1} F(t)>
\nonumber\\
&& +\int\; ds \;\int_{t\le s}\;  dt\;
<e^{-c\omega(s-t)}\cdot F(s),\; \omega^{-1}\cdot F(t)>])
\nonumber\\
&=&
\exp(-\frac{\hbar\kappa}{4}
\int\; ds \;\int\;  dt <e^{-c\omega|s-t|}\cdot F(s),\omega^{-1} F(t)>)
\ea
where in the second step we split the integration domain of the first term and 
expressed the hyperbolic sine in terms of exponentials and in the last 
we combined terms. Note that while $e^{\beta\omega},\; \beta>0$ is not 
defined on Schwarz functions, the actual operator that appears 
in the above integral is $e^{-c|s-t|\omega}$ which is well defined.

The last step consists in the observation that up to a constant 
(\ref{5.19}) just displays the 
integral kernel of the inverse of the operator 
\be \label{5.20}
-[\partial/\partial x^0]^2+\omega^2
\ee
To see this, we note that the Green function of (\ref{5.20}) is (use
translation and reflection invariance $G(x,y)=G(x-y)=G(y-x)$)
\be \label{5.21}
G(x)=\int\; \frac{d^{D+1}k}{(2\pi)^{D+1}} \;
\frac{e^{ik_\mu x^\mu}}{k_0^2+\omega(||\vec{k}||)^2}
\ee
We perform the $k_0$ integral by closing the contour over the infinite 
half circle in the upper/lower complex plane for $x^0>0$ and $x^0<0$ 
respectively. We can then apply the residue theorem and pick up the 
pole contribution at $k_0=\pm i\omega(||\vec{k}||)$ with weight $\pm 2\pi i$
of the remaining holomorphic integrand $e^{ik_0 x^0}/(k_0\pm i 
\omega(||\vec{k}||)$ whence 
\be \label{5.22}
G(x)=\frac{1}{2}\int\; \frac{d^D k}{(2\pi)^D} \; 
e^{-|x^0|\omega(||\vec{k}||)}
\frac{e^{i\vec{k}\cdot \vec{x}}}{\omega(||\vec{k}||)}
=\frac{e^{-|x^0|\omega}}{2\omega}\cdot \delta_x
\ee
Thus, we conclude that the Wiener measure corresponding to our Hamiltonian
parametrised by $\omega$ is a Gaussian measure with zero mean and covariance 
\be \label{5.23}
C:=\hbar\kappa (-\partial_0^2+\omega^2)^{-1}
\ee

\subsection{Hamiltonian formulation from measure}
\label{s5.3}
  
This is standard, see e.g. \cite{GJ87}. We just sketch some of the key 
steps for the sake of completeness. \\
\\
We begin with the Gaussian measure $\mu$ of covariance (\ref{5.23}) 
which is reflection and time translation invariant by inspection and
analyse
the null space of the corresponding reflection positive sesqui-linear 
and symmetric bilinear form. We consider smearing functions of compact,
positive and sharp time support 
\be \label{5.24} 
F(t,x):=\sum_{k=1}^N \; \delta(t,t_k) F_k(x),\; G(t,x):=
\sum_{l=1}^M\; \delta(t,s_l) G_l(x),\;
\ee
with $0<t_1<..<t_N,\; 0<s_1<..<s_M$ and 
$F_k, G_l$ are in ${\cal S}(\mathbb{R}^D)$.
We have with the history Hilbert space ${\cal H}':=L_2(\Gamma,d\mu)$
\ba \label{5.25}
<e^{i\Phi[F]},e^{i\Phi[G]}>_{{\cal H}'}
&=& \mu(e^{i\Phi[G-\theta\cdot F]})
=e^{-\frac{1}{2}<G-\theta\cdot F, 
C\cdot(G-\theta\cdot F)>_{L_2(\mathbb{R}^{D+1},
d^{D+1}x})}
\nonumber\\
&=& e^{-\frac{1}{2} <G,C\cdot G>}\;
e^{-\frac{1}{2} <F,C\cdot F>}\;
e^{\frac{1}{2} <\theta\cdot{F},C\cdot G>+<G,C\cdot \theta\cdot F>}
\ea
Now as we just saw and by reflection invariance of $C$
\ba \label{5.26}
<\theta\cdot F, C\cdot G> &=& <F,C \cdot \theta\cdot G>
=\frac{\hbar\kappa}{2} 
\int\; ds \;\int\;  dt <e^{-c\omega|s-t|}\cdot F(s),\omega^{-1} G(-t)>
\nonumber\\
&=&\frac{\hbar\kappa}{2} 
\int\; ds \;\int\;  dt <e^{-c\omega (s+t)}\cdot F(s),\omega^{-1} G(t)>
\nonumber\\
&=&\frac{\hbar\kappa}{2} 
\sum_{k,l} <e^{-c\omega t_k}\cdot F_k,
\omega^{-1} e^{-c \omega s_l} G_l>_{L_2(\mathbb{R}^D, d^Dx)}
\nonumber\\
&=& <\theta\cdot F, C \hat{G}>
\ea
where in the second step we made use of the positive time support of both
functions and we have defined 
\be \label{5.27}
\hat{G}(t,x):=\delta(t,0) [\sum_l e^{-c \omega s_l} G_l](x)
\ee
Let 
\be \label{5.28}
z:=e^{-\frac{1}{2}[<G,C\cdot G>-<\hat{G},C\cdot \hat{G}>]}
\ee
then 
\be \label{5.29}
<e^{i\Phi[F]},e^{i\Phi[G]}-z e^{i\Phi[\hat{G}]}>_{{\cal H}'}=0
\ee
for all $F$. Thus, we conclude that the function 
$e^{i\Phi[G]}-ze^{i\Phi[\hat{G}]}$ belongs to the null space provided we 
can show that the span of the $e^{i\Phi[F]}$ with $F$ of the form 
(\ref{5.24}) is dense in ${\cal H}'$. To see this, we take any $F$ with 
positive and compact time support in $(0,T]$ and consider the approximant 
\be \label{5.30}
F^N(t):=\sum_{k=1}^{N-1}\; \delta(t,t_k)\;F^N_k(x),\;F^N_k(x):=
\int_{t_k-T/(2N)}^{t_k+T/(2N)}\; dt\; F(t,x)
\ee
with $t_k=kT/N,\;k=1,..,N-1$. It follows
\ba \label{5.30a}
&&||e^{i\Phi[F]}-e^{i\Phi[F^N]}||^2_{{\cal H}'}  
\nonumber\\ 
&=& \mu(e^{i\Phi[F-\theta\cdot F]})+
\mu(e^{i\Phi[F^N-\theta\cdot F^N]})-
\mu(e^{i\Phi[F-\theta\cdot F^N]})-
\mu(e^{i\Phi[F^N-\theta\cdot F]})
\ea
We have for instance
\be \label{5.31}
<F^N, C\cdot F^N> = \frac{\hbar\kappa}{2} 
\sum_{k,l=1}^{N-1} <F^N_k,e^{-c \omega |t_k-t_l|}
\omega^{-1}\; F^N_l>_{L_2(\mathbb{R}^D,d^Dx)}
\ee
which is just a Riemann sum approximation of $<F,C\cdot F>$. The other 
calculations are similar. Thus, the span of these vectors is dense.

It follows that for any $G'$ the vector 
$e^{i\Phi[G']}$ can be approximated arbitrarily well by a vector
$e^{i\Phi[G]}$ with $G$ of the form (\ref{5.24}) and in turn this vector 
is equivalent to a vector with time zero support up to a constant. We 
conclude that the Hilbert space of the theory is the completion 
of the span of vectors with sharp time 
zero support functions $F=\delta(t,0)f$ for which we have 
\be \label{5.32}
<e^{i\Phi[F]},e^{i\Phi[F']}>_{{\cal H}'}=e^{-\frac{\hbar\kappa}{4}
<f,\omega^{-1}f'>}=<e^{i\phi[f]},e^{i\phi[f']}>_{{\cal H}}
\ee 
displaying ${\cal H}=L_2(\mathcal{S}(\mathbb{R}^D),d\nu)$ where $\nu$ is the 
Gaussian measure of covariance $\hbar\kappa/(2\omega)$. Here 
$e^{i\phi[f]}:=[e^{i\Phi[F]}]$ denotes the equivalence class of the 
sharp time zero support vector.
 
To compute the Hamiltonian, we use the definition 
($F(t)=\delta(t,0) f,\;F'(t)=\delta(t,0) f'$)
\ba \label{5.33}
<e^{i\phi[f]},\; e^{-\beta H/\hbar} e^{i\phi[f']}>_{{\cal H}}
&:=& 
< e^{i\Phi[F]},\; R\cdot e^{i\Phi[T_{-\beta}\cdot F']}>
\nonumber\\
&=& 
\mu(e^{i\Phi[T_{\beta}\cdot F'-F]})
\nonumber\\
&=& 
e^{-\frac{1}{2}[<F',C \cdot F'>-<\hat{F}',C\cdot \hat{F}'>]}\;
<e^{i\phi[f]},\; e^{i\phi[\hat{f}']}>_{{\cal H}}
\ea
where $\hat{f}'=e^{-c\beta\omega} \cdot f'$ and $\hat{F}'(t)=\delta(t,0) \hat{f}'$.
As the $e^{i\phi[f]}\Omega,\; \Omega:=1$ span ${\cal H}$ it follows 
\be \label{5.34}
e^{-\beta H/\hbar} e^{i\phi[f]}\Omega
=e^{-\frac{\hbar\kappa}{4}[<f,\omega^{-1}f>
-<e^{-\beta c \omega} f,\omega^{-1}\;e^{-\beta c\omega} f>]}
e^{i\phi[e^{-\beta c \omega }f]} \Omega
\ee
We verify that the contraction (\ref{5.34}) coincides with the 
one as obtained from (\ref{5.3}) and (\ref{5.4}) if $H\Omega=0$. 
We have 
\ba \label{5.35}
&& e^{-\beta H/\hbar} \; a \; e^{\beta H/\hbar}=e^{\beta c \omega}\cdot a
\nonumber\\
&& e^{-\beta H/\hbar} \; a^\ast \; e^{\beta H/\hbar}=e^{-\beta c \omega}\cdot 
a^\ast
\nonumber\\
&& e^{-\beta H/\hbar} \; \phi \; e^{\beta H/\hbar}=
\sqrt{\frac{\hbar\kappa}{2\omega}}\cdot [
e^{\beta c \omega}\cdot a+e^{-\beta c \omega}\cdot a^\ast]
\nonumber\\
&=& e^{-\beta c \omega}\cdot \phi+
\sqrt{\frac{2\hbar\kappa}{\omega}}\;{\rm sh}(\beta c \omega)\cdot a
\nonumber\\
&&e^{-\beta H/\hbar} e^{i\phi[f]}\Omega
=e^{i \phi[e^{-\beta c \omega}\cdot f]+
ia[\sqrt{\frac{2\hbar\kappa}{\omega}}\;{\rm sh}(\beta c \omega)\cdot f]}\;
\Omega
\nonumber\\
&=& e^{\frac{1}{2}[\phi[e^{-\beta c \omega}\cdot f],
a[\sqrt{\frac{2\hbar\kappa}{\omega}}\;{\rm sh}(\beta c \omega)\cdot f]]}\;
e^{i \phi[e^{-\beta c \omega}\cdot f]}\;\Omega
\nonumber\\
&=& 
e^{-\frac{\hbar\kappa}{4}[<f,\omega^{-1} f>
-<e^{-\beta c \omega} f, \omega^{-1} e^{-\beta c \omega} f>]}\;\;
e^{i \phi[e^{-\beta c \omega}\cdot f]}\;\Omega
\ea
indeed.

\section{Discretised theory and renormalisation}
\label{s7}

In this section we simulate the typical situation in constructive QFT 
and pretend not to know what the Hamiltonian or path integral formulation
of the given classical theory should be in the continuum. Hence, we 
will introduce temporal and spatial IR cut-offs $T,R$ respectively and 
work on finite lattices with $M,N$ points in  each spatial or temporal 
direction respectively. We introduce maybe natural but still ad hoc 
discretised versions of the Hamiltonian or the corresponding 
path integral measure 
and apply the both renormalisation procedures derived in \cite{1}. We will determine explicitly 
the corresponding fixed point structure and show that the resulting 
measures or Hamiltonian theories indeed correspond to the continuum theories 
constructed in section \ref{5.1}. \\
\\
The common starting point for both renormalisation trajectories is 
a family of either Gaussian, reflection positive measures $\mu^{(0)}_{R,M}$ 
or equivalently OS data $({\cal H}^{(0)}_{R,M}, \;\Omega^{(0)}_{R,M}, \;
H^{(0}_{R,M})$ originating from some spatial (lattice) discretisation of
the classical continuum theory. In what follows, we construct such a 
discretisation explicitly using a choice of coarse graining map.\\
\\
The fields at finite IR cut-off are supposed to obey periodic boundary 
conditions 
$\phi_R(x+R\delta_a)=\phi_R(x),\; \pi_R(x+R\delta_a)=\pi_R(x)$ for all $\delta_a\in\mathbb{Z}$, 
$a=1,..,D$. The corresponding one particle Hilbert space is 
$L_{R}:=L_2([0,R]^D,d^Dx)$. In the presence of an additional UV cut-off 
we define the one particle Hilbert space as 
$L_{RM}:=\ell_2(\mathbb{Z}_M^D)$ 
with $\mathbb{Z}_M:=\{0,1,..,M-1\}$. These are 
the square summable finite sequences $f_{RM}$ with 
norm squared 
\be \label{5.36}
||f_{RM}||_{L_{RM}}^2:=\epsilon_{RM}^D\; 
\sum_{m\in \mathbb{Z}^D_M} |f_{RM}(m)|^2,\;\;
\epsilon_{RM}:=\frac{R}{M}
\ee 
Here the prefactor $\epsilon_{RM}^D$ is consistent with the interpretation 
that $f_{RM}(m)=f_R(m\epsilon_{RM})$ for {\it some} $f_R\in L_R$ so that 
$||f_{RM}||_{L_{RM}}=||f_R||_{L_R}$. 

Indeed, we have injections 
\be \label{5.36a} 
I_{RM}:\; L_{RM}\to L_{R};\; f_{RM} \mapsto 
\sum_{m\in \mathbb{Z}_M^D} \; f_{RM}(m) \; \chi_{m \epsilon_{RM}}(x)
\ee
with 
\be \label{5.37}
\chi_{m\epsilon_{RM}}(x):=
\prod_{a=1}^D\; \chi_{[m^a\epsilon_{RM},(m^a+1)\epsilon_{RM})}(x)
\ee
These injections are isometric since the $\chi_{m\epsilon_{RM}}$ define a 
partition of $[0,R)^D$ 
\be \label{5.38}
<\chi_{m\epsilon_{RM}},\chi_{m'\epsilon_{RM}}>_{L_{R}}=\epsilon_{RM}^D
\delta_{m m'}
\ee
Likewise, we have evaluation maps (which are densely defined on the  
continuous elements of $L_{R}$)
\be \label{5.39}
E_{RM}:\; L_{R}\to L_{RM};\;f_R \mapsto f_R(m\epsilon_{RM})
\ee 
which satisfy $E_{RM}\circ I_{RM}={\rm id}_{L_{RM}}$.

We consider the discretised fields 
\begin{align} \label{5.40}
\phi_{RM}(m)&:=(I_{RM}^\dagger \phi_R)(m)=\int_{[0,R]^D}\; d^Dx\; 
\chi_{m \epsilon_{RM}}(x)\; \phi_R(x)\nonumber\\
\pi_{RM}(m)&:=(E_{RM}\pi_R)(m):=\pi_R(m\epsilon_{RM}) 
\end{align}
Notice that (\ref{5.40}) defines a partial symplectomorphism 
\be \label{5.41}
\{\pi_{RM}(m),\phi_{RM}(m')\}=\int\; d^Dx \;\chi_{m\epsilon_{RM}}(x)
\;\{\pi_R(x),\phi_R(m'\epsilon_{RM})\}=\chi_{m\epsilon_{RM}}(m'\epsilon_{RM})
=\delta_{m m'}
\ee

Consider the family of Hamiltonians 
\be \label{5.42}
H^{(0)}_{RM}:=\frac{c}{2}\;
\sum_{m \in \mathbb{Z}_M^D}(\kappa\epsilon_{RM}^D
\pi^2_{RM}(m)
+\frac{1}{\epsilon_{RM}^D\kappa} \phi_{RM}(m) 
[(\omega^{(0)}_{RM})^2 \cdot \phi_{RM}](m))
\ee
Here we have defined $\omega^{(0)}_{RM}$ in terms of a suitable, 
self-adjoint (with respect to $L_{RM}$) discretisation $\Delta_{RM}$
of the Laplacian, that is, if the continuum $\omega_R$ is a certain 
function $G=G(-\Delta_R,p^2)$ of the continuum Laplacian $\Delta_R$ on 
$[0,R)^D$ then $\omega^{(0)}_{RM}$ is the function $G(-\Delta_{RM},m^2)$. 
A popular choice is 
\be \label{5.43}
(\Delta_{RM}\cdot f_{RM})(m):=
\frac{1}{\epsilon_{RM}^2}[f_{RM}(m+1)+f_{RM}(m-1)-2f_{RM}(m)]
\ee
It is not difficult to check that (\ref{5.42}) converges to
\be \label{5.44}
H_R:=\frac{c}{2}\int_{[0,R)^D}\; d^Dx\;[\kappa \pi_R^2+\frac{1}{\kappa}
\phi\omega_R^2\phi]
\ee
on smooth fields as $M\to \infty$. 

The form (\ref{5.42}) of the Hamiltonian motivates to introduce discrete 
annihilation operators 
\be \label{5.45}
a^{(0)}_{RM}:=\frac{1}{\sqrt{2\hbar\kappa}}[\sqrt{{\omega^{(0)}_{RM}}/{\epsilon_{RM}^D}}\phi_{RM}
-i\kappa
\sqrt{{\epsilon_{RM}^D}/{\omega^{(0)}_{RM}}}\pi_{RM}]
\ee
so that 
\be \label{5.46}
H^{(0)}_{RM}=\hbar c\sum_{m\in\mathbb{Z}_M} \; (a^{(0)}_{RM})^\ast\; 
\omega^{(0)}_{RM}\cdot a^{(0)}_{RM}
\ee
From (\ref{5.45}) we define a Fock space ${\cal H}^{(0)}_{RM}$ with Fock vacuum 
$\Omega^{(0)}_{RM}$ annihilated by (\ref{5.46}). The Fock space 
can be presented as the Hilbert space ${\cal H}^{(0)}_{RM}=
L_2(\mathbb{R}^{M^D},
d\nu^{(0)}_{RM})$ where $\nu^{(0)}_{RM}$ is a Gaussian measure with covariance
$c_{RM}=\frac{1}{2}\hbar\kappa (\omega^{(0)}_{RM})^{-1}$ with vacuum vector
$\Omega_{RM}=1$. Hence, we have constructed explicitly a family of OS data 
$({\cal H}^{(0)}_{RM},H^{(0)}_{RM},\Omega^{(0)}_{RM})$ which 
certainly is not a fixed point family.

We sidestep the introduction of a temporal cut-off $T$ 
and its corresponding temporal renormalisation and directly
construct the Wiener measure $\mu^{(0)}_{RM}$ on the history spaces 
$\Gamma_{RM}$ of fields $\Phi_{RM}$ corresponding to 
the OS data constructed above. The construction is entirely identical to the 
continuum calculation, hence we know that the Wiener measure $\mu^{(0)}_{RM}$ 
is described by a Gaussian measure 
with the covariance $C^{(0)}_{RM}$ of the operator
\be \label{5.97}
-\frac{1}{c^2}\partial_t^2+[\omega^{(0)}_{RM}]^2
\ee
~\\
Remark:\\
For completeness sake we sketch how one would proceed if one would 
also discretise time. The reader not interested in that aspect can 
skip this remark. If we also discretise time then we would 
also replace (\ref{5.97}) by 
\be \label{5.98}
-\Delta_{TN}+\omega_{RM}^2
\ee
where $\Delta_{TN}$ is a one-dimensional lattice Laplacian defined on 
$\mathbb{Z}_N$, e.g. 
\be \label{5.99}
(\Delta_{TN} F_{TN})(n)=\frac{1}{\delta_{TN}^2}[F_{TN}(n+1)+F_{TN}(n-1)
-2 F_{TN}(n)];\;\;\delta_{TN}=\frac{T}{N}
\ee
where we have introduced a time IR cut-off $T$ and a time resolution $N$. 
In that case we would need coarse graining maps $I_{TNRM}$. 

The isometric injections are respectively given by
\be \label{5.100}
\hat{I}_{RM}:\; L_T\otimes L_{RM}\to L_T\otimes L_R,\;\; 
I_{TN,RM}:\; L_{TN}\otimes L_{RM}\to L_T\otimes L_R
\ee
where $L_T=L_2([0,T], dt)$ and $L_{TN}$ is the space of square summable 
finite sequences in $N$ points with measure $\delta_{TN}=T/N$, i.e. 
\be \label{5.101}
||F_{TN}||_{L_{TN}}^2=\delta_{TN} \sum_{n\in \mathbb{Z}_N}\;
|F_{TN}(n)|^2
\ee
We will take them to be of direct product form 
\be \label{5.101a}
\hat{I}_{RM}=1_{L_T} \otimes I_{RM},\;\;
I_{TN,RM}=I_{TN} \otimes I_{RM}
\ee
where $I_{TN}$ is constructed just as $I_{RM}$ in (\ref{5.36}) with $D=1$
and the substitutions $R\to T, M\to N$. The corresponding evaluation maps 
are 
\be \label{5.100a}
\hat{E}_{RM}:=1_{L_T}\otimes E_{RM}:\; 
L_T\otimes L_R \to L_T \otimes L_{RM},\;\;
E_{TN,RM}:=E_{TN}\otimes I_{RM}:\; L_T \otimes L_R \to L_{TN}\otimes L_{RM}
\ee
and the coarse graining maps are 
\be \label{5.101b}
\hat{I}_{RM\to 2M}=\hat{E}_{R2M}\circ \hat{I}_{R M}=1_{L_T}\otimes I_{RM},\;\;
I_{TN\to 2N,RM\to 2M}=E_{T 2N,R 2M}\circ I_{TN,RM}
\ee
Fix pointing the corresponding measure 
family $\mu_{RM}$ and $\mu_{TN,RM}$ is equivalent to fix pointing the 
corresponding covariances which produces the renormalisation sequences 
\be \label{5.102}
C_{RM}^{(n+1)}=\hat{I}_{RM\to 2M}^\dagger\;C^{(n)}_{R2M}\;
\hat{I}_{RM\to 2M},\;\; 
C_{TN,RM}^{(n+1)}=I_{TN\to 2N,RM\to 2M}^\dagger\;C^{(n)}_{T2N,R2M}\;
I_{TN\to 2NRM\to 2M}
\ee
which is solved by 
\be \label{5.102a}
C_{RM}^\ast=\hat{I}_{RM}^\dagger\;C^\ast\;
\hat{I}_{RM},\;\; 
C_{TN,RM}^\ast=I_{TN,RM}^\dagger\;C^\ast\;
I_{TN,RM}
\ee
for some covariance $C^\ast$ on $L_{TR}$ for instance 
\be \label{5.103}
C^\ast=(-\partial_t^2/c^2+\omega^2)^{-1}
\ee
In our companion papers \cite{6,7} we show that the computations 
that we perform below for free quantum fields in one spatial dimension 
can be extended to any such dimensions. Since, as we will also see below, 
for free quantum fields additional temporal 
renormalisation can be deduced from the purely spatial renormalisation 
of a theory in one more spatial dimension, it follows that we also 
captured the temporal renormalisation scheme which thus leads to 
the fixed point continuum covariance (\ref{5.103}).

\subsection{Path Integral Induced Hamiltonian Renormalisation}
\label{s7.1}

We begin with the path integral induced renormalisation flow.

\subsubsection{Step 1: Computing the path integral flow}
\label{s7.1.1}

Following 
the general programme, the first step
will be to calculate the flow and the fixed points of the measure family 
$\mu^{(0)}_{RM}$. To that end we 
consider the maps 
\be \label{5.47}
I_{R M\to 2M}:=E_{R 2M}\circ I_{RM}:\; L_{RM}\to L_{R 2M}
\ee
which have the property $I_{R 2M} \circ I_{R M\to 2M}=I_{RM}$ as we checked 
in an earlier section. As a 
consequence of this property and the isometry of the maps $I_{RM}$,
(\ref{5.47}) is an isometric injection
\begin{align} \label{5.48}
<I_{R M\to 2M}\cdot f_{RM},I_{R M\to 2M}\cdot f'_{RM}>_{L_{R2M}}&=   
<I_{R 2M}\circ I_{R M\to 2M}\cdot f_{RM},
I_{R 2M}\circ I_{R M\to 2M}\cdot f'_{RM}>_{L_R}
\nonumber\\
= <I_{R M}\cdot f_{RM},
I_{R M} \cdot f'_{RM}>_{L_R}&=<f_{RM},f'_{RM}>_{L_{RM}}
\end{align}
Explicitly for $m\in \mathbb{Z}_{2M}^D$
\be \label{5.48a}
[I_{R M\to 2M} \cdot f_{RM}](m)=\sum_{m'\in \mathbb{Z}_M^D}\;
\chi_{m'\epsilon_{RM}}(m\epsilon_{R2M}) f_{RM}(m')=f_{RM}(\lfloor m/2\rfloor)
\ee
where $\lfloor m/2\rfloor^a:=\lfloor m^a/2\rfloor,\; a=1,..,D$ denotes the component wise Gauss bracket.

The path integral flow is defined by
\be \label{5.48b}
\mu^{(n+1)}_{RM}(e^{i\Phi_{RM}[F_{RM}]}):=
\mu^{(n)}_{R2M}(e^{i\Phi_{R2M}[I_{RM\to 2M}]})
\ee
and it follows immediately that the flow generates a family of Gaussian
measures with covariances $C^{(n)}_{RM}$, since the initial family is such. Namely, we find 
\be \label{5.48c}
C^{(n+1)}_{RM}=(1_{L_2}\otimes I_{RM\to 2M})^\dagger \; C^{(n)}_{R2M} 1_{L_2}\otimes 
I_{R,M\to 2M}
\ee
where the notation is to indicate, that no temporal renormalisation takes 
place.

In the continuum the kernel of the covariance is defined as 
\be \label{5.54}
<F_R,C_R \cdot F_R>_{L_2\otimes L_R}=:
\int_{\mathbb{R}} \; ds\;\int_{\mathbb{R}} \; ds'\;
\int_{[0,R)^D}\; d^Dx\;\int_{[0,R)^D} \; d^Dy\;
F_R(s,x)\; C_R((s,x),(s',y)) \; F_R(s',y)
\ee
It follows 
\ba \label{5.54a}
&& <1_{L_1}\otimes I_{RM} \cdot F_{RM},
C_R \cdot 1_{L_2}\otimes I_{RM} \cdot F_{RM}>_{L_2\otimes L_R}
=\\
&=&<F_{RM},[(1_{L_2} \otimes I_{RM})^\dagger C_R (1_{L_2}\otimes I_{RM})] 
F_{RM}>_{L_2\otimes L_{RM}}
=:<F_{RM},C_{RM} F_{RM}>_{L_{RM}}\nonumber
\ea
which shows that
\be \label{5.54b}
C_{RM}((s,m),(s',m'))=\epsilon_{RM}^{-2D} 
<\chi_{m\epsilon_{RM}},C_R((s,.),(s',.)) \chi_{m'\epsilon_{RM}}>_{L_R}
\ee
Note that the continuum kernel family is automatically a fixed point 
of (\ref{5.48c}) due to $I_{RM}=I_{R 2M}\circ I_{RM \to 2M}$. Expression 
(\ref{5.54b}) tends to $C_R((s,m\epsilon_{RM}),(s',m'\epsilon_{RM}))$ 
as $M\to \infty$.

Explicitly, we have in terms of the kernel of the covariance for the flow 
of the discretised covariance
\begin{align} \label{5.55}
&<F_{RM},C^{(n+1)}_{RM} F_{RM}>_{L_2\otimes L_{RM}}
\\
&= \epsilon_{RM}^{2D}
\sum_{m'_1,m'_2\in \mathbb{Z}_M^D} \int\;ds\;\int\; ds'\;
F_{RM}(s,m'_1) F_{RM}(s',m'_2)
C^{(n+1)}_{RM}((s,m'_1),(s',m'_2))
\nonumber\\
&=
<(1_{L_2}\otimes I_{R M\to 2M})\cdot F_{RM},C^{(n)}_{R2M} 
(1_{L_2}\otimes I_{R M\to 2M})\cdot F_{RM}>_{L_2\otimes L_{R2M}}
\nonumber\\
&= 
\epsilon_{R2M}^{2D}
\sum_{m_1,m_2\in \mathbb{Z}_{2M}^D}\; \int\; ds\int\;ds'\;\times
\nonumber\\
&\;\;\;\;\times (1_{L_2}\otimes  I_{R M\to 2M}\cdot F_{RM})(s,m_1) 
(1_{L_2}\otimes I_{R M\to 2M}\cdot F_{RM})(s',m_2)
C^{(n)}_{R2M}((s,m_1),(s',m_2))
\nonumber\\
&=
\frac{1}{2^{2D}}\epsilon_{RM}^{2D} \; \int\; ds\int\;ds'\;
\sum_{m_1,m_2\in \mathbb{Z}_{2M}^D} 
F_{RM}(\lfloor m_1/2\rfloor ) F_{RM}(\lfloor m_2/2\rfloor)
C^{(n)}_{R2M}((s,m_1),(s',m_2))
\nonumber\\
&=
\frac{1}{2^{2D}}\epsilon_{RM}^{2D} 
\sum_{m'_1,m'_2\in \mathbb{Z}_{M}^D} \; \int\; ds\int\;ds'\;\times \nonumber\\
&\;\;\;\;\times 
F_{RM}(s,m'_1)) F_{RM}(s',m'_2)\sum_{\lfloor m_1/2\rfloor =m'_1,\lfloor m_2/2\rfloor=m_2'}
C^{(n)}_{R2M}((s,m_1),(s',m_2))
\nonumber
\end{align}
from which we read off 
\ba \label{5.57}
C^{(n+1)}_{RM}((s,m'_1),(s',m'_2))&=&2^{-2D}\sum_{\lfloor m_1/2\rfloor =m'_1,\lfloor m_2/2\rfloor =m'_2}
C^{(n)}_{R2M}((s,m_1),(s',m_2))
\nonumber\\
&=& 2^{-2D}\sum_{\delta_1,\delta_2\in \{0,1\}^D} 
C^{(n)}_{R 2M}((s,2m'_1+\delta_1),(s',2m'_2+\delta_2))
\ea
A simplification can be achieved by making use of the translation 
invariance of the (discrete) Laplacian and thus the corresponding covariances
$C_{RM}((s,m),(s',m'))=C_{RM}(s-s',m-m')$, a property which is preserved by 
inspection 
under the block spin transformation (\ref{5.57}). This suggests to 
use Fourier transform techniques.

Recall that $L_R$ is equipped with the orthonormal basis 
$R^{-D/2}e^{ik_R n\cdot x},\; n\in \mathbb{Z}^D,\;x\in [0,R)^D$ where 
$k_R=\frac{2\pi}{R}$. If we restrict $x$ to the lattice points 
$x=m\epsilon_{RM},\;m\in \mathbb{Z}_M^D$ then 
$e^{i k_R n\cdot x}=e^{i k_M n\cdot m},\;k_M=\frac{2\pi}{M}$ and
we may restrict $n$ to $\mathbb{Z}_M^D$ as well. Indeed, we may 
define Fourier transform and its inverse on $L_{RM}$ by 
\be \label{5.58}
f_{RM}(m)=:\sum_{n\in \mathbb{Z}_M^D}\; \hat{f}_{RM}(n) \; e^{i k_M n\cdot m},
\;\;
\hat{f}_{RM}(n)=: M^{-D}\sum_{m\in \mathbb{Z}_M^D}\; f_{RM}(m) \; 
e^{-i k_M n\cdot m}
\ee 
The Fourier transform has the advantage that it diagonalises the Laplacian
$[\Delta_{RM} e^{ik_M n\cdot \;.}](m)=-\hat{\Delta}_{RM}(n k_M) 
e^{ik_M n\cdot m}$
and if $C^{(n)}_{RM}=G(-\partial_t^2,-\Delta_{RM},p^2)$ then 
we have 
\ba \label{5.59}
&& [G\cdot F_{RM}](s,m)
=\epsilon_{RM}^D\sum_{m'\in\mathbb{Z}^D_M} \;\int\;ds'\;G(s-s',m-m') f_{RM}(m')
\\
&=&\sum_{m'\in\mathbb{Z}^D_M}\int ds'\sum_{n\in\mathbb{Z}^D_M}  \int\; \frac{dk_0}{2\pi} \; \hat{F}_{RM}(k_0,n) \;
G(k_0^2,-\hat{\Delta}_{RM}(n k_M),p^2)\; e^{i(k_0 (s-s')+k_M n\cdot (m-m'))}
\nonumber\\
&=& \sum_{m'\in\mathbb{Z}^D_M}\int\; ds'\; F_{RM}(s',m')[M^{-D} \sum_{n\in\mathbb{Z}^D_M} e^{i
(k_0(s-s')+k_M n\cdot (m-m'))} 
G(k_0^2,-\hat{\Delta}_{RM}(n k_M),p^2)]\nonumber
\ea
whence 
\begin{align} \label{5.60}
C_{RM}(s-s',m-m')&=\sum_{n\in\mathbb{Z}^D_M}\;\int\; dk_0\;  e^{i(k_0(s-s')+k_M n\cdot (m-m'))}
\hat{C}_{RM}(k_0,n)\nonumber\\
\hat{C}_{RM}(k_0,n)&=R^{-D}\;
G(k_0^2,-\hat{\Delta}_{RM}(k_M n),p^2)
\end{align}
for the discretised family.

Since for general $\omega_R$ 
it is explicitly only possible to study the flow of the covariance in terms 
of its Fourier transform we translate (\ref{5.57}) in terms of the Fourier 
transform
\ba \label{5.61}
&&C^{(n+1)}_{RM}((s,m'_1),(s',m'_2))= \sum_{l'\in \mathbb{Z}_M^D}\;
\int\;\frac{dk_0}{2\pi}\; 
e^{i (k_0(s-s')+k_M l'\cdot (m-m'))} \;\hat{C}^{(n+1)}_{RM}(k_0,l')
\\
&=& 2^{-2D}\sum_{l\in \mathbb{Z}_{2M}^D}\; \int\;\frac{dk_0}{2\pi}\; 
\hat{C}^{(n)}_{R2M}(k_0,l)
\sum_{\delta_1,\delta_2\in \{0,1\}^D} 
e^{i (k_0(s-s')+k_{2M} l\cdot(2(m'_1-m'_2)+\delta_1-\delta_2))}
\nonumber\\
&=& 2^{-2D}\sum_{l'\in \mathbb{Z}^D_M} \;\int\;\frac{dk_0}{2\pi}\; 
 \sum_{\delta_1,\delta_2,\delta_3\in \{0,1\}^D} \;
\hat{C}^{(n)}_{R2M}(k_0,l'+\delta_3 M)
e^{i (k_0(s-s')+k_{2M} (l'+\delta_3 M)\cdot(2(m'_1-m'_2)+\delta_1-\delta_2))}
\nonumber\\
&=& 2^{-2D}\sum_{l'\in \mathbb{Z}^D_M} e^{ik_M l'\cdot (m'_1-m'_2)}
\;\int\;\frac{dk_0}{2\pi}\; 
\sum_{\delta_1,\delta_2,\delta_3\in \{0,1\}^D} \;
\hat{C}^{(n)}_{R2M}(l'+\delta_3 M)
e^{i (k_0(s-s')+k_{2M} (l'+\delta_3 M)\cdot(\delta_1-\delta_2))}
\nonumber
\ea
whence 
\be \label{5.62}
\hat{C}^{(n+1)}_{RM}(k_0,l')=2^{-2D}
\sum_{\delta_1,\delta_2,\delta_3\in \{0,1\}^D} \;
\hat{C}^{(n)}_{R2M}(k_0,l'+\delta_3 M)
e^{i k_{2M} (l'+\delta_3 M)\cdot(\delta_1-\delta_2)}
\ee
~\\
\\
We will now carry out the details of this procedure for illustrative 
purposes for the case $D=1$ and the Poincar\'e invariant choice
\be \label{5.62a}
\omega_R=\sqrt{-\Delta_R+p^2}     
\ee
More general models in all dimensions will be discussed in our companion 
papers. 
For $D=1$ (\ref{5.62}) becomes with $l'\in \mathbb{Z}_M$
\ba \label{5.63}
\hat{C}^{(n+1)}_{RM}(k_0,l')
&=&\frac{1}{2}
\sum_{\delta_3\in \{0,1\}} \;
\hat{C}^{(n)}_{R2M}(k_0,l'+\delta_3 M)
[1+\cos(k_{2M} (l'+\delta_3 M))]
\\
&=&\frac{1}{2}\{
\hat{C}^{(n)}_{R 2M}(k_0,l')[1+\cos(k_{2M} l')]
+\hat{C}^{(n)}_{R 2M}(k_0,l'+M)[1-\cos(k_{2M} l')]\}\nonumber
\ea
We start the flow with $\hat{C}^{(0)}_{RM}(k_0,l'):=\hat{C}_{RM}(k_0,l')$ 
where 
$C_{RM}(k_0, l')$ corresponds to the naive discretisation of the Laplacian 
(\ref{5.43}). Thus from (\ref{5.60}) with $l\in \mathbb{Z}_M$
\be \label{5.63a}
\hat{C}^{(0)}_{RM}(k_0,l)=R^{-1} \frac{\hbar\kappa}{2} 
\frac{1}{2\epsilon_{RM}^{-2}[1-\cos(k_M l)]+k_0^2+p^2}
\ee
It is equivalent to study the flow of 
$\hat{c}_{RM}(l):=2R\hat{C}_{RM}(l)/(\hbar\kappa)$ and it is convenient 
to introduce the abbreviations $t:=k_M l, q:=\sqrt{k_0^2+p^2}\epsilon_{RM}$. 
Hence 
\be \label{5.64}
\hat{c}^{(0)}_{RM}(l)=
\frac{\epsilon_{RM}^2}{2[1-\cos(t)]+q^2}
\ee
For reasons that will become transparent in a moment we rewrite (\ref{5.64})
as follows: Let 
\be \label{5.65}
a_0(q):=1+q^2/2,\; b_0(q):=q^3/2, c_0(q):=0
\ee
then trivially
\be \label{5.66}
\hat{c}^{(0)}_{RM}(l)=\frac{\epsilon_{RM}^2}{q^3}\;
\frac{b_0(q)+c_0(q) \cos(t)}{a_0(q)-\cos(t)}
\ee
The purpose of doing this trivial rewriting is that it turns out that 
the parametrisation by 3 functions $a_n,b_n,c_n$ of $q$ in the Ansatz
\be \label{5.67}
\hat{c}^{(n)}_{RM}(l)=\frac{\epsilon_{RM}^2}{q^3}\;
\frac{b_n(q)+c_n(q) \cos(t)}{a_n(q)-\cos(t)}
\ee
is invariant under the renormalisation flow. Namely by (\ref{5.63})
(note $t=k_{M}l\to k_{2M} l=t/2, q=\sqrt{k_0^2+p^2}\epsilon_{RM}\to 
\sqrt{k_0^2+p^2}\epsilon_{R 2M}=q/2$
and $\cos(k_{2M}(l+M))=-\cos(t/2)$)
\ba \label{5.68}
&& \hat{c}^{(n+1)}_{RM}(l) =\frac{\epsilon_{RM}^2}{q^3}\;
\frac{b_{n+1}(q)+c_{n+1}(q) \cos(t)}{a_{n+1}(q)-\cos(t)}
\\
&=&
\frac{1}{2}\{
\hat{c}^{(n)}_{R 2M}(l)[1+\cos(k_{2M} l)]
+\hat{c}^{(n)}_{R 2M}(l+M)[1-\cos(k_{2M} l)]\}
\nonumber\\ 
&=& \frac{\epsilon_{R 2M}^2}{2 (q/2)^3}
\{\frac{b_n(q/2)+c_n(q/2)\cos(t/2)}{a_n(q/2)-\cos(t/2)}[1+\cos(t/2)]
+\frac{b_n(q/2)-c_n(q/2)\cos(t/2)}{a_n(q/2)+\cos(t/2)}[1-\cos(t/2)]\}
\nonumber\\
&=& \frac{\epsilon_{RM}^2}{q^3} \frac{1}{a_n(q/2)^2-\cos^2(t/2)}
\{[b_n(q/2)+c_n(q/2)\cos(t/2)][a_n(q/2)+\cos(t/2)][1+\cos(t)]
\nonumber\\
&& +[b_n(q/2)-c_n(q/2)\cos(t/2)][a_n(q/2)-\cos(t/2)][1-\cos(t/2)]\}
\nonumber\\
&=& \frac{\epsilon_{RM}^2}{q^3} 
\frac{1}{a_n(q/2)^2-\frac{1}{2}[1+\cos(t/2)]}
\{2 a_n(q/2) b_n(q/2)+2\cos^2(t/2)[b_n(q/2)+c_n(q/2)+a_n(q/2) c_n(q/2)]\}
\nonumber\\
&=& \frac{\epsilon_{RM}^2}{q^3} 
\frac{1}{[2a_n(q/2)^2-1]-\cos(t)}
\{2[2 a_n b_n+b_n+c_n+a_n c_n](q/2)+2[b_n+c_n+a_n c_n](q/2)\cos(t)\}
\nonumber
\ea
We deduce the recursion relations 
\ba \label{5.69}
a_{n+1}(q) &=& 2 a_n(q/2)^2-1
\nonumber\\
b_{n+1}(q) &=& 2[2 a_n b_n+b_n+c_n+a_n c_n](q/2)
\nonumber\\
c_{n+1}(q) &=& 2[b_n+c_n+a_n c_n](q/2)
\ea
The corresponding fixed point equations become coupled functional equations 
\ba \label{5.70}
a_\ast(q) &=& 2 a_\ast(q/2)^2-1
\nonumber\\
b_\ast(q) &=& 2[2 a_\ast b_\ast+b_\ast+c_\ast+a_\ast c_\ast](q/2)
\nonumber\\
c_\ast(q) &=& 2[b_\ast+c_\ast+a_\ast c_\ast](q/2)
\nonumber
\ea
The easiest of these three equations is the first one, as it involves only 
one function and we easily recognise the functional equation of the
cosine or hyperbolic cosine. Now, $a_0(q)>1$ for $q>0$ and assuming 
this to be the case also for $a_n(q)$ we get $a_{n+1}(q)=2 a_n(q/2)^2-1
>1$ for $q>0$. It follows $a_\ast(q)>1$ for  $q>0$ so that 
\be \label{5.71}
a_\ast(q)={\rm ch}(q)
\ee
Next we observe 
\be \label{5.72}
d_\ast(q):=(b_\ast+c_\ast)(q)=4[1+a_\ast(q/2)][b_\ast+c_\ast](q/2)
\ee
which is a homogeneous linear functional equation 
as $a_\ast$ is already known. If we 
define $[b_\ast+c_\ast](q)=q^n P({\rm ch}(q))$, where $P$ is a polynomial,
then we have a chance to satisfy the fixed point equation since $q^n$ 
can take the factor of 4 into account and the rhs depends only on ch$(q/2)$
as well as the lhs as ch$(q)=2$ch$^2(q/2)-1$. Let then 
$P=\sum_{k=0}^N z_k {\rm ch}^k(q)$ then we find in terms of $x={\rm ch}(q/2)$ 
\be \label{5.73}
2^n\sum_{k=0}^N z_k [2 x^2-1]^k=
4(x+1)\sum_{k=0}^N z_k x^k
=4\{z_0+z_N x^{N+1}+\sum_{k=1}^N [z_k+z_{k-1}] x^k\}
\ee
We may assume that $z_N\not=0$, otherwise decrease the degree of the 
polynomial. Then we must have $2N=N+1$ i.e. $N=1$. It follows 
\be \label{5.74}
2^n\{z_0-z_1+2 z_1 x^2\}=4\{z_0+(z_0+z_1)x+z_1 x^2\}
\ee
i.e.
\be \label{5.75}
n=1,\;z_1=-z_0 \;\Rightarrow \;\; z q({\rm ch}(q)-1)
=:d_\ast(q)
\ee
where $z$ is a constant to be determined later. 

Finally, we have
\be \label{5.76}
c_\ast(q)=2(b+c)_\ast(q/2)+2 a_\ast(q/2) c_\ast(q/2)=2 d_\ast(q/2)+
2 a_\ast(q/2) c_\ast(q/2)
\ee
which is an inhomogeneous linear functional equation as $a_\ast,d_\ast$
are already known. The general solution will therefore be the linear combination
of a special solution $c_1$ of the inhomogeneous equation and the general 
solution $c_2$ of the corresponding homogeneous equation. Explicitly
\be \label{5.77}
0=-c_1(q)+z q({\rm ch}(q/2)-1)+2{\rm ch}(q/2)c_1(q/2)
=-[c_1(q)+zq]+{\rm ch}(q/2)[2 c_1(q/2)+zq]
\ee
which is solved by $c_1(q)=-z q$. This leaves us with 
\be \label{5.78}
c_2(q)=2 {\rm ch}(q/2) c_2(q/2)
\ee
which is the functional equation of $c_2(q)=z' {\rm sh}(q)$ where again $z'$
is a constant to be determined later. 

To see which values $z,z'$ are chosen by the initial functions of the 
fixed point equation we notice that $d_0(q)=q^3/2$ and assume 
$\lim_{q\to 0} 2 d_n(q)/q^3=1$ up to some $n$ then also 
\be \label{5.79}
\lim_{q\to 0} \frac{2 d_{n+1}(q)}{q^3}
=\lim_{q\to 0} \frac{8[{\rm ch}(q/2) +1]d_n(q/2)}{q^3}
=\lim_{q\to 0} \frac{2 d_n(q/2)}{(q/2)^3}=1
\ee
Thus also 
$\lim_{q\to 0} 2 d_\ast(q)/q^3=1$ whence $z=1$. Finally, we have 
$c_0(q)=0$ hence $\lim_{q\to 0} c_0(q)/q^3$ is regular. We assume this 
to be the case up to some $n$, i.e. $c_n(q)=O(q^3)$. Then 
\be \label{5.79b}
c_{n+1}(q)/q^3=2 d_n(q/2)/q^3+ 2 a_n(q/2) c_n(q/2)/q^3
\ee
is also regular at $q=0$ hence so must be $c_\ast(q)$. It follows $z'=1$.

We summarise: The fixed point equation is uniquely solved by 
\ba \label{5.79a}
a_\ast(q) &=& {\rm ch}(q)
\nonumber\\
b_\ast(q) &=& q {\rm ch}(q) - {\rm sh}(q)
\nonumber\\
c_\ast(q) &=& {\rm sh}(q)- q
\ea
~\\
We now compare this to the known continuum theory. The continuum theory 
is described by the covariance $C_R=\frac{\hbar\kappa}{2} 
(-\partial_t^2-\Delta_R+p^2)^{-1}$
or equivalently $c_R=\frac{2R}{\hbar\kappa} C_R=
R(-\partial_t^2-\Delta_R+p^2)^{-1}$
which can now be directly compared to (\ref{5.64}). The corresponding
cylindrical projection at resolution $M$ is  
\ba \label{5.80}
c_{RM}((s,m),(s',m')) &=& \epsilon_{RM}^{-2} 
([1_{L_2}\otimes I_{RM})^\dagger c_R [1_{L_2}\otimes I_{RM}])((s,m),(s',m'))
\nonumber\\
&=&\epsilon_{RM}^{-2} 
\int_{m\epsilon_{RM}}^{(m+1)\epsilon_{RM}}\; dx\;
\int_{m'\epsilon_{RM}}^{(m'+1)\epsilon_{RM}}\; dy\; c_R((s,x),(s',y))
\ea
where $c_R(x,y)$ is the continuum kernel. To compute it we employ again 
Fourier transformation and use the fact that the functions $e_{nR}(x)
=e^{in k_R x}/\sqrt{R},\;k_R=2\pi/R$ form an orthonormal basis on 
$L_R=L_2([0,R),dx)$. Hence 
\ba \label{5.81}
c_R((s,x),(s',y)) &=& R (-\partial_s^2-\Delta_{Rx}+p^2)^{-1} 
\delta_{\mathbb{R}}(s,s')\delta_R(x,y) 
\\
&=&\int \frac{dk_0}{2\pi}\;
\sum_{n\in\mathbb{Z}} \; e_{-nR}(y)\; R (-\partial_s^2-\Delta_{Rx}+p^2)^{-1} 
e^{i k_0(s-s')}\;e_{nR}(x)\nonumber\\
&=& \int \frac{dk_0}{2\pi}\;\sum_{n\in\mathbb{Z}} \; 
\frac{e^{i (k_0(s-s')+k_R n(x-y))}}{(n k_R)^2+k_0^2+p^2}
\nonumber
\ea
It follows 
\ba \label{5.82a}
&& c_{RM}((s,m),(s',m')) =\\
&=&\epsilon_{RM}^{-2} 
\sum_{n\in\mathbb{Z}} \;\int \frac{dk_0}{2\pi}\;
\frac{e^{ik_0(s-s')}}{(n k_R)^2+k_0^2+p^2}\;
[\int_{m\epsilon_{RM}}^{(m+1)\epsilon_{RM}}\; dx\; e^{i k_R n x}]
[\int_{m'\epsilon_{RM}}^{(m'+1)\epsilon_{RM}}\; dy\; e^{i k_R n y}]^\ast
\nonumber\\
&=&
\epsilon_{RM}^{-2} 
\sum_{n\in\mathbb{Z}} \; \int \frac{dk_0}{2\pi}\;
\frac{e^{ik_0(s-s')}}{(n k_R)^2+k_0^2p^2}\;
[\epsilon_{RM}\delta_{n,0}+\frac{1-\delta_{n,0}}{i k_R n} (
e^{i k_R n (m+1)\epsilon_{RM}}-e^{i k_R n m\epsilon_{RM}})]\times
\nonumber\\
&&\;\times[\epsilon_{RM}\delta_{n,0}
-\frac{1-\delta_{n,0}}{i k_R n} (
e^{-i k_R n (m'+1)\epsilon_{RM}}-e^{-i k_R n m'\epsilon_{RM}})]
\nonumber\\
&=&
\sum_{n\in\mathbb{Z}} \;
\int \frac{dk_0}{2\pi}\; 
\frac{e^{i (k_0(s-s')+k_M n (m-m')}}{(n k_R)^2+k_0^2+p^2}\;
[\delta_{n,0}+2\frac{1-\delta_{n,0}}{(k_M n)^2}(1-\cos(k_M n)]\nonumber
\ea
To compare this expression to $\hat{c}_{RM}(k_0,l),\; l\in \mathbb{Z}_M$ 
we write $n=l+NM,\; N\in \mathbb{Z}$ and split the sum
\begin{align} \label{5.82}
&c_{RM}((s,m),(s',m')) =\\
&=
\sum_{l\in \mathbb{Z}_M}\;
\int \frac{dk_0}{2\pi}\;
e^{i (k_0(s-s')+k_M l (m-m')}\; 
\sum_{N\in\mathbb{Z}} \; 
\frac{1}{([l+NM] k_R)^2+k_0^2+p^2}\;\frac{2(1-\cos([l+NM]k_M)}{([l+NM] k_M)^2}\nonumber
\end{align}
where we declare the last fraction to equal unity at $l=N=0$.
Comparing with the first line of (\ref{5.61}) we see that the sum
involved in (\ref{5.61}) coincides with the definition of 
$\hat{c}_{RM}(k_0,l)$ .

We now carry out the sum over $N$ by employing the Poisson resummation formula
\be \label{5.83}
\sum_{N\in \mathbb{Z}} f(N)=\sum_{N\in \mathbb{Z}} \int_\mathbb{R} 
e^{-2i \pi Nx}  f(x)
\ee
with 
\be \label{5.84}
f(x):=
\frac{1}{([l+xM] k_R)^2+k_0^2+p^2}\;\frac{2(1-\cos([l+xM]k_M)}{([l+xM] k_M)^2}
\ee
to which the Poisson resummation may be applied as $f$ is smooth and decays 
at infinity as $1/x^4$. We find with $q=\sqrt{k_0^2+p^2} \epsilon_{RM}$ 
\ba \label{5.85}
\hat{c}^\infty_{RM}(k_0,l) &=& \sum_{N\in\mathbb{Z}} \; 
\frac{1}{([l+NM] k_R)^2+k_0^2+p^2}\;\frac{2(1-\cos([l+NM]k_M))}{([l+NM] k_M)^2}
\nonumber\\
&=& \epsilon_{RM}^2 
\sum_{N\in\mathbb{Z}} \;\int \; dx \; e^{-i 2\pi N x}\; 
\frac{1}{([l+xM] k_M)^2+q^2}\;\frac{2(1-\cos([l+xM]k_M))}{([l+xM] k_M)^2}
\nonumber\\
&=& \epsilon_{RM}^2 
\sum_{N\in\mathbb{Z}} \;\int \; dx \; e^{-i 2\pi N x}\; 
\frac{1}{(k_Ml+2\pi x)^2+q^2}\;\frac{2(1-\cos(k_M l+2\pi x))}{(k_M l+2\pi x)^2}
\nonumber\\
&=& \epsilon_{RM}^2 
\sum_{N\in\mathbb{Z}} \;\int \; \frac{dx}{2\pi} \; e^{-i N x}\; 
\frac{1}{(k_Ml+x)^2+q^2}\;\frac{2(1-\cos(k_M l+x))}{(k_M l+x)^2}
\nonumber\\
&=& \epsilon_{RM}^2 
\sum_{N\in\mathbb{Z}} \; e^{i k_M N l}\;
\int \; \frac{dx}{2\pi} \; \frac{2 e^{-i N x}(1-\cos(x))}{x^2(x^2+q^2)}
\ea
We have 
\be \label{5.86}
\frac{2(1-\cos(x))e^{-iNx}}{x^2}=
\frac{e^{-iNx}-e^{-i(N-1)x}}{x^2}+\frac{e^{-iNx}-e^{-i(N+1)x}}{x^2}
\ee
For any $N$ (\ref{5.86}) 
is holomorphic in the entire complex plane and for $N\ge 1$ 
decays on the lower infinite 
half-circle and  for $N\le -1$ 
decays on the upper infinite half-circle. It follows
that the integrand is holomorphic everywhere in the whole complex plane 
except at $x=\pm i q$ and the contour can be closed as described for 
$N\not=0$. Thus we find by the 
residue theorem
\ba \label{5.87}    
&&
\int \; \frac{dx}{2\pi} \; \frac{2 e^{-i N x}(1-\cos(x))}{x^2(x^2+q^2)}
\nonumber\\
&=&
\left\{ \begin{array}{cc}
-\frac{2\pi i}{2\pi}\frac{2 e^{-i N x}(1-\cos(x))}{x^2(x-iq)}_{x=-iq}
& N\ge 1 \\
\frac{2\pi i}{2\pi}\frac{2 e^{-i N x}(1-\cos(x))}{x^2(x+iq)}_{x=iq}
& N\le -1 
\end{array}
\right.
\nonumber\\
&=&
\frac{e^{-|N|q}({\rm ch}(q)-1)}{q^3}
\ea
For $N=0$ the first term in (\ref{5.86}) decays on the 
upper while the second decays on the lower infinite half circle. However,
the two terms are not separately holomorphic at $x=0$, only their sum is.
We thus write the integral as a principal value integral $\lim_\delta\to 0$ 
where we leave out the interval $[-\delta,\delta]$ and then
close the contour for the first/second term with a small half-circle of radius 
$\delta$ in the upper/lower complex plane, subtract that added contribution 
and apply the residue theorem. We find  
\ba \label{5.88}    
&&
\int \; \frac{dx}{2\pi} \; \frac{2(1-\cos(x))}{x^2(x^2+q^2)^2}
\nonumber\\
&=&
-\frac{2\pi i}{2\pi}\frac{(1-e^{-ix})}{x^2(x-iq)}_{x=-iq}
-\lim_{\delta\to 0} \int_{x=\delta e^{i\phi},\;\phi\in[-\pi,0]}\;
\frac{dx}{2\pi x}\; 
\frac{1-e^{-ix}}{x}\;\frac{1}{x^2+q^2} 
\nonumber\\
&& +
\frac{2\pi i}{2\pi}\frac{(1-e^{ix})}{x^2(x+iq)}_{x=iq}
+\lim_{\delta\to 0} \int_{x=\delta e^{i\phi},\;\phi\in[0,\pi]}\;
\frac{dx}{2\pi x}\; 
\frac{1-e^{ix}}{x}\;\frac{1}{x^2+q^2} 
\nonumber\\
&=& 
\frac{q+e^{-q}-1}{q^3}
\ea
It remains to compute the geometric sum in (\ref{5.85}) with $k_M l=t$
\ba \label{5.89}
\frac{\hat{c}_{RM}(l)}{\epsilon_{RM}^2}
&=&
\frac{q+e^{-q}-1}{q^3}
+\frac{{\rm ch}(q)-1}{q^3}(-2+\sum_{N=0}^\infty\;
\{e^{-N[q+i t]}+e^{-N[q-it]} \})
\nonumber\\
&=&
\frac{q+e^{-q}-1-{\rm ch}(q)+1}{q^3}
+\frac{{\rm ch}(q)-1}{q^3}(-1
+\frac{1}{1-e^{-[q+it]}}
+\frac{1}{1-e^{-[q-it]}})
\nonumber\\
&=&
\frac{q-{\rm sh}(q)}{q^3}
+\frac{{\rm ch}(q)-1}{q^3}\;\frac{2-2 e^{-q}\cos(t)-
[1+e^{-2q}-2 e^{-q} \cos(t)]}{1+e^{-2q}-2 e^{-q} \cos(t)}
\nonumber\\
&=&
\frac{q-{\rm sh}(q)}{q^3}
+\frac{{\rm ch}(q)-1}{q^3}\;\frac{{\rm sh}(q)}{{\rm ch}(q)-\cos(t)}
\nonumber\\
&=& 
\frac{1}{q^3}\; \frac{1}{{\rm ch}(q)-\cos(t)}\;
\{[{\rm ch}(q)-1]{\rm sh}(q)+[q-{\rm sh}(q)]{\rm ch}(q)-[q-{\rm sh}(q)]
\cos(t)\}
\nonumber\\
&=& 
\frac{1}{q^3}\; \frac{1}{{\rm ch}(q)-\cos(t)}\;
\{[q {\rm ch}(q) - {\rm sh}(q)]+[{\rm sh}(q)-q]\cos(t)\}
\ea
Comparing with (\ref{5.67}) and (\ref{5.79a}) we see that we obtain 
{\it perfect match}! The fixed point equations of the naively
discretised covariance of the history field measure 
have found the {\it precise} cylindrical projections of 
its continuum covariance $[-\partial_t^2-\Delta_R+p^2]^{-1}$. It is therefore 
clear that the fixed point of the path integral induced Hamiltonian renormalisation precisely delivers the continuum 
OS data via OS reconstruction that we started from and that we artificially
discretised. Moreover, it is easy to see that the continuum 
limit $\lim_{M\to \infty} c_{RM}(l)=[k_0^2+m^2+(lk_R)^2]^{-1}$ coincides with 
the continuum covariance.

\subsubsection{Step 2: 
Computing the path integral induced Hamiltonians}
\label{s7.1.2}

We carry out the explicit OS reconstruction of the measures $\mu^{(n)}_R$.
To do this recall that these are determined 
by the covariances of their Fourier transform which up to a factor 
follow the flow ($l\in \mathbb{Z}_M$), see (\ref{5.63})
\ba \label{5.200}
\hat{c}^{(n+1)}_{R,M}(k_0,l) &=&   
\frac{1}{2}\{
\hat{c}^{(n)}_{R 2M}(k_0,l)[1+\cos(t/2)]
+\hat{c}^{(n)}_{R 2M}(k_0,l+M)[1-\cos(t/2)]\},\;
\nonumber\\
\hat{c}^{(0)}_{R M}(k_0,l) &=&\frac{\epsilon_{RM}^2}{2[1-\cos(t)]+q^2},\;\;
t=k_M l,\;\; q=\sqrt{p^2+k_0^2}\epsilon_{RM}
\ea
We conclude that while $c^{(0)}_{RM}(k_0,l)$ displays only one simple pole
with respect to $q^2$,
the number of poles gets doubled at each renormalisation step. For instance
\be \label{5.201}
\hat{c}^{(1)}_{RM}(k_0,l)
=\frac{\epsilon_{RM}^2}{8}\{
\frac{1+\cos(t/2)}{2[1-\cos(t/2)]+q^2/4}
+\frac{1-\cos(t/2)}{2[1+\cos(t/2)]+q^2/4}
\}
\ee
It follows that $c^{(n)}_{R,M}(k_0,l)$ displays $2^n$ distinct simple 
poles in $q^2$. Thus, in the parametrisation 
\be \label{5.202}
\hat{c}^{(n)}(k_0,l)=\frac{\epsilon_{RM}^2}{q^3}
\frac{b_n(q)+c_n(q) \cos(t)}{a_n(q)-\cos(t)}
\ee
$a_n(q)$ must be a polynomial of order $2^n$ in $q^2$ whose poles can 
in principle be 
read off from the flow equation (\ref{5.200}). These poles are certain 
mutually distinct functions of $t$ as one easily sees from the inductive 
definition (\ref{5.200}). The flow equation (\ref{5.200})
for the covariance at resolution $M$ is a superposition of covariances 
at resolution $2M$ with merely $t$ dependent, positive coefficients 
$1\pm \cos(t/2)$
(i.e. they do not depend on $q$). Since $c^{(0)}_{RM}(k_0,t)$ has 
its $q$ dependence only in the pole and a single positive coefficient, 
this feature is preserved for the entire flow.
We may therefore display an alternative 
parametrisation of (\ref{5.202}) as follows: Let 
$-[\hat{\lambda}^{(n)}_{RMN}(t)]^2$
with $N=-2^{n-1}+1,-2^{n-1}+2,..,0, 1,.., 2^{n-1}$ denote the poles of 
$c^{(n)}_{RM}(k_0,l)$ for $n>0$ with respect to $q^2$ (from the induction
(\ref{5.200}) it follows that the poles are strictly non-positive) and 
$\hat{g}^{(n)}_{RMN}(t)\epsilon_{RM}^2$ the corresponding, positive 
coefficient functions. 
Then for $n>0$
\be \label{5.203}
\hat{c}^{(n)}_{RM}(k_0,l)=\sum_{N=-2^{n-1}+1}^{2^{n-1}} 
\frac{\hat{g}^{(n)}_{RMN}(t)\epsilon_{RM}^2}{q^2
+[\hat{\lambda}^{(n)}_{RMN}(t)]^2}
\ee
If we trivially extend $\hat{g}^{(n)}_{RMN}\equiv 0$ 
for $N>2^{n-1},N\le -2^{n-1}$ we may 
extend the sum over $N$ to infinity
\be \label{5.204}
\hat{c}^{(n)}_{RM}(k_0,l)=\sum_{N\in\mathbb{Z}} 
\frac{\hat{g}^{(n)}_{RMN}(t)\epsilon_{RM}^2}{q^2
+[\hat{\lambda}^{(n)}_{RMN}(t)]^2}
\ee
which now provides a universal parametrisation for the $c^{(n)}_{R}(k_0,t)$.
The flow is now in terms of the poles and their respective coefficient 
functions. More and more coefficient functions are switched on from
zero to a positive function as the flow number $n$ increases. We even know
what the fixed point values of this flow are, if we look at (\ref{5.82})
\be \label{5.205}
[\hat{\lambda}^\ast_{RMN}(t)]^2=(t+2\pi N)^2,\;\hat{g}^\ast_{RMN}(t)
=2\frac{1-\cos(t)}{(t+2\pi N)^2}
\ee

The first question is, what null space of the reflection positive 
inner product for a covariance of the form (\ref{5.204}) results.       
It is convenient to introduce the renormalisation invariant lattice Laplacian 
\be \label{5.206}
(\Delta_{RM} f_{RM})(m):=f_{RM}(m+1)+f_{RM}(m-1)-2f_{RM}(m)
\ee
on $L_{RM}$ 
which in Fourier space corresponds to multiplication by 
$2(\cos(k_M l)-1)=2(\cos(t)-1)$. The functions $\hat{\lambda}^{(n)}_{RMN}(t),
\;\hat{g}^{(n)}_{RMN}(t)$ can now be considered as the eigenvalues 
of corresponding operator valued functions of $\Delta_{RM}$ (using the 
spectral theorem in the form $t=\pm {\rm arcos}(cos(t))$ for $0<t<2\pi$ and 
noticing that the flow generates a function which is invariant under 
$t\to -t$ so that the sign ambiguity is irrelevant) which we denote 
by $\lambda^{(n)}_{RMN},g^{(n)}_{RMN}$ respectively. We also set 
\be \label{5.207}
[\omega^{(n)}_{RMN}]^2:=\frac{[\lambda^{(n)}_{RMN}]^2}{\epsilon_{RM}^2} + 
p^2 
\ee
Then we find for the corresponding reflection positive inner product 
for functions $F_{RM}, G_{RM}$ of positive time support (we drop the label 
$n$ for a moment)
\ba \label{5.208}
&& <[e^{i\Phi_{RM}[F_{RM}]}]_{\mu_{RM}},
[e^{i\Phi_{RM}[G_{RM}]}]_{\mu_{RM}}>_{\mu_{RM}}
=\mu_{RM}(e^{i\Phi_{RM}[\theta\cdot G_{RM}-F_{RM}]})
\\
&=& 
\mu_{RM}(e^{i\Phi_{RM}[F_{RM}]})\;
\mu_{RM}(e^{i\Phi_{RM}[G_{RM}]})\;
\times \nonumber\\ &&
\exp(-\sum_N\int\; ds\int ds' \int \frac{dk_0}{2\pi} \; 
e^{i k_0(s-s')}\;
<F_{RM}(s),\frac{g_{RMN}}{k_0^2+\omega_{RMN}^2}\; 
G(-s')_{RM}>_{L_{RM}})
\nonumber\\
&=& 
\mu_{RM}(e^{i\Phi_{RM}[F_{RM}]})\;
\mu_{RM}(e^{i\Phi_{RM}[G_{RM}]})\;
\times \nonumber\\ &&
\exp(-\sum_N\int\; ds\int ds' \int \frac{dk_0}{2\pi} \; 
e^{i k_0(s+s')}\;
<F_{RM}(s),\frac{g_{RMN}}{k_0^2+\omega_{RMN}^2}\; 
G(s')_{RM}>_{L_{RM}})
\nonumber\\
&=& 
\mu_{RM}(e^{i\Phi_{RM}[F_{RM}]})\;
\mu_{RM}(e^{i\Phi_{RM}[G_{RM}]})\;
\times \nonumber\\ &&
\exp(-\sum_N\int\; ds\int ds' \; 
<F_{RM}(s),\frac{g_{RMN} \; e^{-(s+s')\omega_{RMN}}}{2\omega_{RMN}}\; 
G(s')_{RM}>_{L_{RM}})
\nonumber
\ea
We now extract representatives of the equivalence classes
of the inner product (\ref{5.208}) corresponding to  
fields not at a single sharp time zero, but rather a countably infinite set 
of sharp times. To see how this comes about, we compute, noticing the 
time support of $G$ and dropping all labels for the sake of the argument 
\ba \label{5.209}
&&\int\; ds \; e^{-s\omega} G(s,m)
=\sum_l\;\int\; \frac{dk_0}{2\pi}\;\int_0^\infty ds\;
e^{i (k_0 s+k_M l m)} e^{-s \omega(l)} \hat{G}(k_0,l)   
\nonumber\\
&=& \sum_l\;\int\; \frac{dk_0}{2\pi}\;\frac{1}{\omega(l)-i k_0} 
e^{i k_M l m} \hat{G}(k_0,l)   
=-\sum_l e^{i k_M l m} \hat{G}(k_0=-i\omega(l),l)
\ea
by the residue theorem. Here we used that $G(s)=0$ for $s<0$ implies that 
its Fourier transform $\hat{G}(k_0)$ is holomorphic on the lower complex half
plane with at most polynomial growth at infinity. Hence the residue theorem 
applies. It follows that the value of (\ref{5.209}) remains unchanged 
if we replace $\hat{G}(k_0,l)$ by $\hat{G}'(k_0,l)=
f(k_0,l)\;\hat{G}(-i\omega(l),l)$ where
$h$ is a fixed function holomorphic in the lower half plane such that 
$h(-i\omega(l),l)=1$, e.g. $h\equiv 1$. 

If we have a finite number of frequencies $\omega_N>0$ labelled by 
$N\in \mathbb{Z}$ then likewise we consider the functions 
$\hat{G}_N(l):=\hat{G}(-i\omega_N(l),l)$ and can replace $\hat{G}(k_0,l)$ 
by 
\be \label{5.210}
\hat{G}'(k_0,l)=\sum_N\; h_N(k_0,l)\; \hat{G}_N(l),\; h_N(-i\omega_{N'}(l),l)
=\delta_{N,N'}
\ee
A possible choice is 
\be \label{5.211a}
h_N(k_0,l):=\prod_{N'\not= N}\; 
\frac{e^{-i\tau k_0}-e^{-\tau\omega_{N'}(l)}}   
{e^{-\tau\omega_N(l)}-e^{-\tau\omega_{N'}(l)}}   
\ee
where $\tau>0$ is any fixed positive real number. (\ref{5.211a}) is well 
defined because the pole values $\omega_N(l)$ are mutually distinct for 
different $N$ and equal $l$. It is holomorphic everywhere and a polynomial
in $e^{-ik_0 \tau}$ where the order coincides with the number of different
frequencies $\omega_N$ reduced by one, in our case this number is given 
by $2^n-1$. Thus it becomes a constant at the 
lower half circle in the complex plane of infinite radius and the residue
theorem applies. We conclude that the function $G'(s,m)$ itself, at the 
n-th renormalisation step, has the form 
\be \label{5.211}
G'(s,l)=\sum_{r=0}^{2^n-1} \; \delta(s-r\tau)\; g_r(l), \; g_r\in L_{RM}
\ee
i.e. they depend on $2^n$ sharp points of time rather than a single one,
except for $n=0$! It follows that $[e^{i\Phi_{RM}[G_{RM}]}]_{\mu^{(n)}_{RM}}$
can be identified with the representative 
\be \label{5.212}
\frac{\mu^{(n)}_{RM}(e^{i\Phi_{RM}[G_{RM}]})}
{\mu^{(n)}_{RM}(e^{i\Phi_{RM}[G'_{RM}]})}\;
e^{i\Phi_{RM}[G'_{RM}]} 
\ee
or in other words 
\be \label{5.213}
e^{i\Phi_{RM}[G_{RM}]}-\frac{\mu^{(n)}_{RM}(e^{i\Phi_{RM}[G_{RM}]})}
{\mu^{(n)}_{RM}(e^{i\Phi_{RM}[G'_{RM}]})}\;
e^{i\Phi_{RM}[G'_{RM}]} 
\ee
is a null vector with respect to the reflection positive inner product 
defined by $\mu^{(n)}_{RM}$. The OS Hilbert space ${\cal H}^{(n)}_{RM}$ 
can thus be thought of as 
the completion of the finite linear span of the $e^{i\Phi_{RM}[G'_{RM}]}$ with 
$G'_{RM}$ of the form (\ref{5.211}) and $\Omega^{(n)}_{RM}\equiv 1$. 

We compute the corresponding Hamiltonian. This amounts to computing 
the representative of the equivalence class of 
$e^{i\Phi_{RM}[T_\beta \cdot F_{RM}]}$ for $F_{RM}$ of the form 
(\ref{5.211}). We have 
\be \label{5.214}
(T_\beta\cdot F_{RM})(s)=F_{RM}(s-\beta)=
\sum_r\; \delta(s-\beta,r\tau)\; f^r_{RM}
=\sum_r\; \delta(s,\beta+r\tau)\; f^r_{RM}
\ee
whence 
\be \label{5.215}
\widehat{T_\beta\cdot F_{RM}}(k_0,l)=\sum_r\; 
e^{-ik_0(\beta+r\tau)}\;\hat{f}^r_{RM}(l)
\ee
Thus 
\begin{align} \label{5.216}
\widehat{T_\beta\cdot F_{RM}}'(k_0,l)
&=\sum_N \; h_{RMN}(k_0,l)\; 
\widehat{T_\beta\cdot F_{RM}}'(-i\omega_{RMN}(l),l)
=\nonumber\\
&=\sum_N \; h_{RMN}(k_0,l)\;\sum_r\;  
e^{-\omega_{RMN}(l)(\beta+\tau r)}\;\hat{f}^r_{RM}(l)
\end{align}
where $h_{RMN}$ is defined as in (\ref{5.211}) with $\omega_N$ replaced by 
$\omega_{RMN}$ and we suppressed the renormalisation step label $n$ for 
notational convenience. If we decompose 
\be \label{5.217}
h_{RMN}(k_0,l)=:\sum_r\; e^{-ir\tau k_0}\; h^r_{RMN}(l)
\ee
we obtain 
\be \label{5.218}
\widehat{T_\beta\cdot F_{RM}}'(k_0,l)
=\sum_r e^{-i k_0 \tau r} \; \sum_{r'}\;  
[\sum_N h^r_{RMN}(l)\; e^{-\omega_{RMN}(l)(\beta+\tau r')}]\;\
\hat{f}^r_{RM}(l)
\ee
Accordingly, the time evolution is described by the matrix 
\be \label{5.219}
A^{r,r'}_{RM}(\beta,l):=
\sum_N h^r_{RMN}(l)\; e^{-\omega_{RMN}(l)(\beta+\tau r')}
\ee
or in position space by the corresponding matrix valued operator 
where $\omega_{RMN}(l)$ is replaced by the corresponding operator.
It follows that we can describe the time translation contraction semigroup
on the chosen representatives, reintroducing the renormalisation step label, by 
\be \label{5.220}
e^{-\beta H^{(n)}_{RM}} e^{i\Phi_{RM}[F_{RM}]}
=\frac{\mu^{(n)}_{RM}(e^{i\Phi_{RM}[F_{RM}]})}
{\mu^{(n)}_{RM}(e^{i\Phi_{RM}[A^{(n)}_{RM}(\beta)\cdot F_{RM}]})}
\; e^{i\Phi_{RM}[A^{(n)}_{RM}(\beta)\cdot F_{RM}]},\;
F_{RM}=\sum_r \delta_{r\tau} f^r_{RM}
\ee
where $A^{(n)}_{RM}(\beta)$ is the purely spatial matrix valued operator 
whose Fourier transform is displayed in (\ref{5.219}). It is instructive
to verify the semigroup law
\be \label{5.221}
A^{(n)}_{RM}(\beta_1)\cdot A^{(n)}_{RM}(\beta_2)
=A^{(n)}_{RM}(\beta_1+\beta_2)
\ee
which rests on the van der Monde identity for polynomials of degree 
$d=2^{n}-1$
\be \label{5.222}
p(x)=\sum_{r=0}^d\; a_r\; x^r,\;\;p(x_r)=p_r,\; x_0< x_1<..<x_d\;\;
\Rightarrow\;\; p(x)=\sum_{r=0}^d\; 
p_r\;\prod_{r'\not=r}\;\frac{x-x_{r'}}{x_r-x_{r'}}
\ee
We prove it by applying (\ref{5.210}) to $h_{RMN'}(\omega_{RMN}(l)\tau,l)$ in
\begin{align}
&\sum_{r'}A^{r,r'}_{RM}(\beta_1,l)A^{r',r''}_{RM}(\beta_2,l)=\nonumber\\
&=\sum_{NN'}h^r_{RMN}(l) [\sum_{r'}e^{-\omega_{RMN}(l)\tau r'} h^{r'}_{RMN'}(l)]e^{-\omega_{RMN'}(l)\tau r''}e^{-\omega_{RMN}(l)\beta_1-\omega_{RMN'}(l)\beta_2}=\nonumber\\
&=
\sum_{NN'}h^r_{RMN}(l) \delta_{N,N'}e^{-\omega_{RMN'}(l)\tau r''}e^{-\omega_{RMN}(l)\beta_1-\omega_{RMN'}(l)\beta_2}=A^{r,r''}_{RM}(\beta_1+\beta_2)
\end{align}
As $n\to \infty$ and for fixed finite $M$, the Hilbert space can thus no 
longer be thought of as described by a single sharp time zero field but 
rather by
sharp time fields at an exponentially increasing (with $n$) number of 
sharp times. At the fixed point thus, the number of this sharp points of time 
is actually infinite. How can this be reconciled with the fact that in the 
continuum the Hilbert space {\it can} be described by a single field 
at sharp time zero? The answer lies in the continuum limit $M\to \infty$:
If we inspect (\ref{5.205}) then we see that at fixed $l\in \mathbb{Z}_M$ 
we obtain for the coupling ``constants'' $\hat{g}_{RMN}(l)\to \delta_{N,0}$
as $M\to \infty$. At the same time $\omega_{RMN}(l)$ 
diverges for all $N$ except $N=0$ and the time contraction for all modes 
except for $N=0$ ``freezes''. Thus in the continuum limit, the theory 
is described by a single dispersion relation and thus the single sharp time 
zero description that we are used to applies.  

The description using fields at more than one sharp time that we have arrived
at means that we cannot express the Hamiltonian in terms of a single time 
zero field and its conjugate momentum. Thus our discussion suggests to 
introduce instead an infinite number of sharp time zero field species 
$\phi_{RNM}$ and their conjugate momenta $\pi_{RMN}$, that is, the 
non-vanishing commutators are 
\be \label{5.223}
{[}\pi_{RMN}(m),\phi_{RMN'}(m')]=i\hbar\delta_{N,N'}\delta_{m,m'},\;
m,m'\in \mathbb{Z}_M
\ee
At finite $n$ of course we only have $N\in\{-2^{n-1}+1,..,0,..,2^{n-1}$\}, i.e. 
we have only $d=2^n$ field species. Accordingly, instead of $L_{RM}=l_2(M)$ 
we consider $L_{RM}=l_2(M)^d$ as the one particle Hilbert space and the
Hamiltonian
\ba \label{5.224}
H'_{RM} &:=& \frac{1}{2}\sum_{(m,N),(m',N')}\;
[\pi_{RMN}(m) D_{RM}((m,N),(m',N')) \pi_{RMN'}(m')
\nonumber\\
&& + \phi_{RMN}(m) E_{RM}((m,N),(m',N')) \phi_{RMN'}(m')]
\nonumber\\
&=:&\frac{1}{2}
[<\pi_{RM},D_{RM} \pi_{RM}>_{L_{RM}}+<\phi_{RM},E_{RM} \phi_{RM}>_{L_{RM}}]
\ea
for certain operators $D_{RM},E_{RM}$ on $L_{RM}$. 
Then we claim that it is possible to choose  
$D_{RM}, E_{RM}$ such that the Wiener measure corresponding to 
(\ref{5.224}) reproduces the path integral measure. To see this, we drop all 
labels for simplicity
\be \label{5.225}
H=\frac{1}{2}[<\pi,D\pi>+<\phi,E\phi>]
\ee
where $D,E$ are self-adjoint, positive and symmetric on $L_{RM}$ and in general 
not commuting. We define annihilators and frequency 
\be \label{5.226}
a=\frac{1}{\sqrt{2}}[<\kappa,\phi>-i<\kappa^{-1},\pi>],\;
:H:=<a^\ast,\omega' a>
\ee
where normal ordering is with respect to $a$. Note that $\kappa,\omega'$ 
are operators on $L_{RM}$. This leads to the identities
\be \label{5.227}
\kappa^\dagger \omega'\kappa=E,\; (\kappa^{-1})^\dagger \omega'\kappa^{-1}=D
\ee
which are solved by 
\be \label{5.228}
\kappa=\kappa^\dagger>0,\; \kappa
=\sqrt{ E^{1/2} \sqrt{ E^{-1/2} D^{-1} E^{-1/2}} E^{1/2}},\;
\omega'=(\omega')^\dagger>0,\;
\omega'=\kappa D \kappa
\ee
Now a simple computation similar to the one for the continuum in section
\ref{s5.1} that generalises the choice $\kappa=\sqrt{\omega'}$ 
shows that the Wiener measure corresponding to 
(\ref{5.226}) yields 
\be \label{5.229}
\mu(e^{i\Phi[F]})=e^{-\frac{1}{2}\int\; ds\;\int\;ds'
<F(s),\frac{e^{-|s-s'|\omega'}}{2\kappa^2}\; F(s')>}
\ee
We now pick the sharp time zero Weyl elements to 
be 
\be \label{5.231}
w_{RM}[f'_{RM}]:=e^{i\sum_N \phi_{RMN}[f'_{RMN}]},\;f'_{RM}=\{f'_{RMN}\}_N\in 
L_{RM}^{2^n}
\ee
and also $W_{RM}[F'_{RM}]=\prod_N\; W_{RMN}[F'_{RMN}],\; 
W_{RMN}[F'_{RMN}]=e^{i\Phi_{RMN}[F'_{RMN}]}$.
Then 
the corresponding Wiener measure gives 
\be \label{5.232}
\mu'_{RM}(W_{RM}[F'_{RM}])
=\exp(-\frac{1}{2}\sum_N 
\int ds\; \int ds'\;<F'_{RM}(s),\frac{e^{-|s-s'|\omega'_{RM}}}{2\kappa^2_{RM}} 
F'_{RM}(s')>
)
\ee
We can use our knowledge from the continuum theory to infer that the 
Hilbert space corresponding to the reflection positive inner product 
of $\mu'_{RM}$ is labelled by time zero smearing functions 
$F'_{RM}(s)=\delta(s,0) f'_{RM}$ and that the Hamiltonian is defined by 
\be \label{5.233}
e^{-\beta H'_{RM}} e^{i\phi_{RM}[f'_{RM}]}
=\frac{\mu'_{RM}(e^{\Phi_{RM}[\delta_0 f'_{RM}]})}
{\mu'_{RM}(e^{\Phi_{RM}[\delta_0 e^{-\beta\omega'_{RM}} f'_{RM}])})}
\;e^{i\phi_{RM}[e^{-\beta\omega'_{RM}} f'_{RM}])}
\ee

To match this to (\ref{5.220}) we perform a trivial relabelling 
between $r,r',\in \{0,...,d-1\}$ and $N,N'\in \{-2^{n-1}+1,..,2^{n-1}\}$
in order to write the matrix elements of $A_{RM}$ in the form 
$A_{RM}((m,N),(m',N');\beta)$. Then the semigroup property (\ref{5.221}) 
implies that there exists a positive self-adjoint generator $\omega_{RM}$
on $L_{RM}$, such that $A_{RM}(\beta)=e^{-\beta \omega_{RM}}$. Next, for 
\be \label{5.234}
F_{RM}=\sum_{r=0}^{d-1}  \delta_{r\tau} f^r_{RM}=:\sum_{N=-2^{n-1}+1}^{2^{n-1}}
\delta_{(N+2^{n-1}-1)\tau}\; f_{RMN}
\ee
we find a positive matrix $B_{RM}$ on $L_{RM}$ such that 
\be \label{5.235}
\mu_{RM}(e^{i \Phi_{RM}[F_{RM}]})=
e^{-\frac{1}{4}<f_{RM},B_{RM} f_{RM}>_{L_{RM}}} 
\ee
If we now compare (\ref{5.220}), (\ref{5.235}) and (\ref{5.232}), (\ref{5.233})
we see that we obtain perfect match provided that we pick
\be \label{5.236}
\omega'_{RM}:=\omega_{RM},\;\; \kappa_{RM}^{-2}:=B_{RM}
\ee

Accordingly, the path integral induced Hamiltonian theory does have an 
interpretation in terms of sharp zero-time fields, however, at the price 
of introducing more and more field species at each renormalisation step.
These field species are mutually commuting, however, the Hamiltonian
couples them to each other according to the matrices $D_{RM},E_{RM}$ 
constructed above.

\subsection{Direct Hamiltonian Renormalisation}
\label{s7.2}

We now discuss the direct Hamiltonian renormalisation in terms of the 
single canonical field species $\phi_{RM}$ at sharp time zero.

\subsubsection{Step 1: Implementing Isotropy}
\label{s7.2.1}

As already remarked before, implementing isotropy of 
\be \label{5.242}
j^{(n)}_{RM\to 2M} e^{i\phi_{RM}[f_{RM}]}\Omega^{(n+1)}_{RM}
:=e^{i\phi_{R2M}[I_{RM\to 2M}\cdot f_{RM}]}\Omega^{(n)}_{R2M}
\ee
is equivalent to studying the flow of the family of Hilbert space measures
\be \label{5.243}
\nu^{(n+1)}_{RM}(e^{i\phi_{RM}[f_{RM}]}):=
\nu^{(n)}_{R2M}(e^{i\phi_{R2M}[I_{RM\to 2M}\cdot f_{RM}]})
\ee
Again it is clear that the family stays Gaussian if the original family is.
Let $\frac{1}{2 \omega^{(0)}_{RM}}$ be the covariance of $\nu^{(0)}_{RM}$.
We have the basic identity (in the sense of the spectral theorem)
\be \label{5.244}
\frac{1}{2 \omega^{(0)}_{RM}}=\int\; \frac{dk_0}{2\pi} 
\frac{1}{k_0^2+(\omega^{(0)}_{RM})^2}
\ee
which, as in the previous section, can be written in terms of 
$q^2=(p^2+k_0^2)\epsilon_{RM}^2$ and $\Delta_{RM}$ (or $t=k_M l,\;
l\in \mathbb{Z}_M$ when Fourier transforming). 
We now make the self-consistent assumption that the 
covariance of $\nu^{(n)}_{RM}$ can also be written in the form
\be \label{5.245}
\frac{1}{2 \omega^{(n)}_{RM}}=
\int\; \frac{dk_0}{2\pi} \; c^{(n)}(q,\Delta_{RM})
\ee
If we compare this to (\ref{5.64}) then we see that the work has already been
done in the previous subsection. Namely, precisely 
the flow of $\omega^{(n)}_{RM}$ 
has been computed there,
the difference with the current section is that 
we restrict the smearing fields to the special time dependence
$F_{RM}=\delta_0\; f_{RM}$. The integral over $k_0$ could be explicitly carried
out using the residue theorem, in particular for the fixed point covariance 
in the form displayed in (\ref{5.82}). However, for what follows we do 
not need to do this.

\subsubsection{Step 2: Computing the direct Hamiltonian flow}

The fixed point sequence 
is defined by the matrix element equations 
\ba \label{5.92}
&&<e^{i<\phi_{RM},f_{RM}>}\Omega^{(n+1)}_{RM},\;H^{(n+1)}_{RM} 
\;e^{i<\phi_{RM},f'_{RM}>}\Omega^{(n+1)}_{RM}>_{{\cal H}^{(n+1)}_{RM}}
\nonumber\\
&:=&
<e^{i<\phi_{R2M},I_{RM\to 2M}\cdot f_{RM}>}\Omega^{(n)}_{R2M},\;H^{(n)}_{R2M} 
\;e^{i<\phi_{R2M},I_{RM\to 2M}\cdot f'_{RM}>}
\Omega^{(n)}_{R2M}>_{{\cal H}^{(n)}_{R2M}}
\ea
Let us define 
\be \label{5.92a}
a_{RM}^{(n)}:=\frac{1}{\sqrt{2}}[
\sqrt{\omega^{(n)}_{RM}}\cdot \phi_{RM}
-i\sqrt{\omega^{(n)}_{RM}}^{-1}\cdot \pi_{RM}]
\ee 
Then 
\be \label{5.92b}
\nu^{(n)}_{RM}(e^{i\phi_{RM}[f_{RM}]})=
e^{-\frac{1}{4}<f_{RM},(\omega^{(n)}_{RM})^{-1}\;f_{RM}>}=<\Omega^{(n)}_{RM},\;
e^{i\phi_{RM}[f_{RM}]}\Omega^{(n)}_{RM}>
\ee
is the Fock measure labelled by (\ref{5.245}) and 
$\Omega^{(n)}_{RM}$ is the Fock vacuum annihilated by (\ref{5.92a}).
Then 
\ba \label{5.93}
&&e^{-i<\phi_{R2M},I_{RM\to 2M}\cdot f'_{RM}>}\;
a^{(n)}_{R2M}(m)\;
e^{i<\phi_{R2M},I_{RM\to 2M}\cdot f'_{RM}>}
\nonumber\\
&=& a^{(n)}_{R2M}(m)-i[<\phi_{R2M},I_{RM\to 2M}\cdot f'_{RM}>,a_{RM}(m)]
\nonumber\\
&=& a_{R2M}(m)
+i\sqrt{\frac{\hbar\kappa \epsilon_{R2M}^{D/2}}{2}}
([\omega^{(n)}_{R2M}]^{-1/2} I_{R M\to 2M}
\cdot f_{RM})(m) 
\ea
We now prove by induction that 
\be \label{5.93b}
H^{(n)}_{RM}=<a^{(n)}_{RM},\omega^{(n)}_{RM}\cdot a^{(n)}_{RM}>_{L_{RM}}
\ee
which is consistent with $H^{(n)}_{RM}\Omega^{(n)}_{RM}=0$.
By construction, (\ref{5.93b}) holds for $n=0$ and all $M$ and 
we assume it to hold up to $n$ and all $M$. Then  
\ba \label{5.94}
&& <e^{i<\phi_{R2M},I_{RM\to 2M}\cdot f_{RM}>}\Omega^{(n)}_{R2M},\;
H^{(n)}_{R2M} 
\;e^{i<\phi_{R2M},I_{RM\to 2M}\cdot f'_{RM}>}
\Omega^{(n)}_{R2M}>_{{\cal H}^{(n)}_{R2M}}
\nonumber\\
&=& \epsilon^D_{RM} \sum_{m,m'}\; \omega^{(n)}_{R2M}(m,m')\;
<a^{(n)}_{R2M}\; e^{i<\phi_{R2M},I_{RM\to 2M}\cdot f_{RM}>}\Omega^{(n)}_{R2M},\;
a^{(n)}_{R2M}
\;e^{i<\phi_{R2M},I_{RM\to 2M}\cdot f'_{RM}>}
\Omega^{(n)}_{R2M}>_{{\cal H}^{(n)}_{R2M}}
\nonumber\\
&=& \frac{\hbar^2 \kappa c}{2}\; \sum_{m,m'}\; \omega^{(n)}_{R2M}(m,m')\;
\overline{([\omega^{(n)}_{R2M}]^{-1/2}\cdot I_{R M\to 2M}\cdot f_{RM})(m)}\;
([\omega^{(n)}_{R2M}]^{-1/2}\cdot I_{R M\to 2M}\cdot f'_{RM})(m'),
\;\;\times
\nonumber\\
&&\;\;\times <e^{i<\phi_{R2M},I_{RM\to 2M}\cdot f_{RM}>}\Omega^{(n)}_{R2M}, 
\;e^{i<\phi_{R2M},I_{RM\to 2M}\cdot f'_{RM}>}
\Omega^{(n)}_{R2M}>_{{\cal H}^{(n)}_{R2M}}
\nonumber\\
&=& \frac{\hbar^2 \kappa c}{2}\; 
<I_{R M\to 2M}\cdot f_{RM},\;I_{R M\to 2M}\cdot f'_{RM}>_{L_{R2M}}
\;\;\times
\nonumber\\
&&\;\;\times <e^{i<\phi_{R2M},I_{RM\to 2M}\cdot f_{RM}>}\Omega^{(n)}_{R2M}, 
\;e^{i<\phi_{R2M},I_{RM\to 2M}\cdot f'_{RM}>}
\Omega^{(n)}_{R2M}>_{{\cal H}^{(n)}_{R2M}}
\nonumber\\
&=& \frac{\hbar^2 \kappa c}{2}\; 
<f_{RM},\;f'_{RM}>_{L_{RM}}
\;\;\times
\nonumber\\
&& <e^{i<\phi_{RM},f_{RM}>}\Omega^{(n+1)}_{RM}, 
\;e^{i<\phi_{RM},f'_{RM}>}
\Omega^{(n+1)}_{RM}>_{{\cal H}^{(n+1)}_{RM}}
\nonumber\\
&=:&
<e^{i<\phi_{RM},f_{RM}>}\Omega^{(n+1)}_{RM},\;H^{(n+1)}_{RM} 
\;e^{i<\phi_{RM},f'_{RM}>}\Omega^{(n+1)}_{RM}>_{{\cal H}^{(n+1)}_{RM}}
\ea
where we have made use of isometry of both $I_{RM\to 2M}$ and 
$J^{(n)}_{RM\to 2M}$.
Thus the matrix elements of $H^{(n+1)}_{RM}$ are consistent with 
(\ref{5.93b}). The fixed point Hamiltonian $H^\ast_{RM}$ is then simply
(\ref{5.93b}) with $\omega^{(n)}_{RM}$ replaced by $\omega^\ast_{RM}$.

We claim that 
\be \label{5.95}
H^\ast_{RM}=J_{RM}^\dagger H_R J_{RM}
\ee
where $H_R$ is the continuum Hamiltonian and $J_{RM}:\; {\cal H}^\ast_{RM} \to 
{\cal H}^\ast_R$ the isometric embedding of Fock spaces which is granted to 
exist
due to the equivalence of the fixed point family to an inductive limit 
Hilbert space family. Indeed in our case this is simply given by 
\be \label{5.96}
J_{RM} e^{i<\phi_{RM},f_{RM}>_{L_{RM}}}\;\Omega^\ast_{RM}=
e^{i<\phi,I_{RM}\cdot f_{RM}>_{L_R}}\; \Omega^\ast_R
\ee
This follows because the isometry check and $J_{R 2M} J_{RM \to 2M}=J_{RM}$
are equivalent to the corresponding statements for $I_{RM}, I_{RM\to 2M}$ and 
to the statement 
$(\omega^\ast_{RM})^{-1}=I_{RM}^\dagger (\omega^\ast_R)^{-1} I_{RM}$ for 
the fixed point covariances of the Hilbert space measures. 
To prove (\ref{5.95})  
we compute, using the same steps as in (\ref{5.94})
\ba \label{5.96a}
&&<e^{i<\phi_{RM},f_{RM}>}\Omega^\ast_{RM},\; [J_{RM}^\dagger H_R J_{RM}]\;
\;e^{i<\phi_{RM},f'_{RM}>}\;\Omega^\ast_{RM}>_{{\cal H}^*_{RM}}
\nonumber\\
&=& 
<e^{i<\phi_R,I_{RM}\cdot f_{RM}>}\Omega^\ast_R,\; H_R \;
\;e^{i<\phi_R,I_{RM}\cdot f'_{RM}>}\;\Omega^\ast_R>_{{\cal H}_R^\ast}
\nonumber\\
&=& \frac{\hbar^2 \kappa c}{2}\; 
<I_{RM}\cdot f_{RM},I_{RM} \cdot f'_{RM}>_{L_R}\;
<e^{i<\phi,I_{RM}\cdot f_{RM}>}\Omega^\ast_R,  \;
\;e^{i<\phi,I_{RM}\cdot f'_{RM}>}\;\Omega^\ast_R>_{{\cal H}_R^\ast}
\nonumber\\
&=& \frac{\hbar^2 \kappa c}{2}\; 
<f_{RM},\;f'_{RM}>_{L_{RM}}\;
<e^{i<\phi_{RM},f_{RM}>}\Omega^\ast_{RM},\;
\;e^{i<\phi_{RM},f'_{RM}>}\;\Omega^\ast_{RM}>_{{\cal H}_{RM}^\ast}
\nonumber\\
&=& <e^{i<\phi_{RM},f_{RM}>}\Omega^\ast_{RM},\; H_{RM} \;
\;e^{i<\phi_{RM},f'_{RM}>}\;\Omega^\ast_{RM}>_{{\cal H}^*_{RM}}
\ea
as claimed. Thus, not only does there exist a consistent family of 
Hamiltonian quadratic forms but indeed a fixed point Hamiltonian $H^\ast_R$.
This Hamiltonian coincides with the one $H_R$ of the continuum because 
we checked in the previous subsection that the fixed point covariances
$\omega^\ast_{RM}$ are obtained from the continuum covariance $\omega_{RM}$
by $[\omega^\ast_{RM}]^{-1}=I_{RM}^\dagger \omega_R^{-1} I_{RM}$. 

It is instructive to check that 
\be \label{5.96b}
\omega_R^{-1}=\lim_{M\to \infty} (\omega^\ast_{RM})^{-1}
\ee
by using the explicit presentation (\ref{5.89}). To do this note that 
with $q^2=(k_0^2+p^2)\epsilon_{RM}^2$ and $t=k_M l$ we have 
$\frac{1}{q^3}[q{\rm ch}(q)-\rm{sh}(q)]\to 1$ as $M\to \infty$ and 
$\frac{1}{q^3}[{\rm sh}(q)-q]\to 1$ as $M\to \infty$ and $\cos(t)\to 1$ and 
$\frac{\epsilon_{RM}^2}{{\rm ch}(q)-\cos(t)}\to 
\frac{1}{k_0^2+p^2+(k_R l)^2}$. Note also that (\ref{5.82}) is an 
instance of the theorem of Mittag-Leffler applied to (\ref{5.89}) 
that allows to write a meromorphic function as a linear combination 
of simple pole functions and an entire holomorphic function.

\subsection{Comparison of the Renormalisation Flows}
\label{s7.3}

The two flows corresponding to the path integral induced Hamiltonian 
renormalisation and the direct Hamiltonian renormalisation of the 
OS data are very different
even for free fields: While the path integral flow generates an infinite 
number of field species at every finite resolution $M$, even if one starts 
with a single field species, the direct flow stays in the single field 
species regime for every finite resolution $M$. The corresponding OS data 
are different and if one would compute the Wiener measure corresponding
to the direct flow OS data we cannot possibly obtain the finite resolution 
path integral measure since even the one particle Hilbert spaces are 
very different from each other. But this is not what the direct Hamiltonian 
flow is 
supposed to do. Rather, its aim is to construct the continuum Hamiltonian
$H_R$ as the limit $\lim_{M\to \infty} H^\ast_{RM}$ where $\{H^\ast_{RM}\}_M$
is the fixed point family as obtained from the direct flow. From these 
continuum OS data we can construct the corresponding Wiener measure and 
then compute its cylindrical projections. As we have seen, it is those 
cylindrical projections which do agree with the finite resolution 
path integral measures at the fixed point of the path integral induced 
flow, at least for the present free field theory that we considered.

\section{Summary}
\label{s8}

In this paper we tested our proposal for a direct Hamiltonian 
renormalisation flow on the OS data for the case of a free Klein
Gordon field in two spacetime dimensions. Generalisations to more dimensions 
and more general models will be supplied in \cite{6,7}. We find that 
the flow has a fixed point whose corresponding continuum Hamiltonian 
agrees with the continuum Fock space quantisation of the classical
continuum Hamiltonian. That this is not only a fixed point but that 
the initial naively discretised family in fact converges to 
it will be demonstrated in \cite{6}. 

The direct flow has the advantage that 
one can sidestep the construction of the Wiener measures, that is, the 
corresponding path integrals which has obvious practical implications.
However, as the present paper reveals, there is in fact a more important 
implication: The direct Hamiltonian flow is much closer to immediately
constructing the finite resolution matrix elements of the actual continuum 
Hamiltonian. By contrast, the flow of the OS data induced by the 
path integral flow is very far away from that, it leads to 
finite resolution Hamiltonians 
with an ever-increasing number of field species as we increase the 
number of renormalisation steps that the continuum theory 
does not have. 

How can these two facts be reconciled? After all, the continuum Hamiltonian,
which only uses a single sharp time zero field species, gives rise to a 
path integral Wiener measure, which therefore also uses only a single 
spacetime field species. The finite resolution 
cylindrical projections of that continuum
measure, which is an OS measure by construction \cite{1}, are also OS 
measures and thus have OS data. Yet, these OS data involve the  
inflation of field species derived in this paper. The reason lies 
in the fact that the OS construction enforces that the contraction 
semigroups of the continuum measure and its cylindrical projections 
are equivariant with respect to the corresponding Hilbert space 
embedding. This implies that the finite resolution Hamiltonians must 
commute with a huge number of subprojection operators. This enhanced 
symmetry of the finite resolution Hamiltonians cannot be accommodated by
a single field species and leads to their inflation.

It is clear that this enhanced symmetry is an unphysical property 
of the finite resolution matrix elements of the continuum Hamiltonian.
In other words, the path integral induced flow of the finite resolution OS
data is physically meaningless. The only meaningful OS data 
that the path integral flow extracts are the OS data of the fixed point
continuum measure. On the other hand, the direct renormalisation flow 
of the OS data runs into a fixed point which does display the finite 
resolution matrix elements of the continuum Hamiltonian and thus 
the flow directly generates approximations to those matrix elements 
that improve their continuum properties eventually with each 
renormalisation step.\\
\\
\\
{\bf\large Acknowledgements}\\
Part of this work was financially supported by a grant from the 
Friedrich-Alexander University to the Emerging Fields Project ``Quantum
Geometry'' under its Emerging Fields Initiative. 
K. L. thanks the German National Merit Foundation for financial support.  
T.L. thanks the Heinrich-B\"oll Foundation for financial support.


\begin{thebibliography}{99}

\parskip -5pt



\bibitem{1} Thorsten Lang, Klaus Liegener, Thomas Thiemann. Hamiltonian Renormalisation I. Derivation from Osterwalder-Schrader Reconstruction.
{\it Class.Quant.Grav.} {\bf 35} 24, 245011 (2018)
[arXiv:1711.05685]

\bibitem{6} Thorsten Lang, Klaus Liegener, Thomas Thiemann. 
Hamiltonian Renormalisation III.
Renormalisation Flow of 1+1 dimensional free, scalar quantum 
of Free Scalar Fields: Properties. 
{\it Class.Quant.Grav.} {\bf 35} 24, 245013 (2018)
[arXiv:1711.05688]

\bibitem{7} Thorsten Lang, Klaus Liegener, Thomas Thiemann.
Hamiltonian Renormalisation IV.
Renormalisation Flow of D+1 dimensional free, scalar quantum 
of Free Scalar Fields and Rotation Invariance. 
{\it Class.Quant.Grav.} {\bf 35} 24, 245014 (2018)
[arXiv:1711.05695]


  \bibitem{GJ87}
 J.~Glimm and A.~Jaffe,
  ``Quantum Physics - A functional integral Point of View'',
  Springer-Verlag, New York, (1987)
 

\bibitem{OS72}
 K.~Osterwalder and R.~Schrader,
  ``Axioms for Euclidean Green's Functions'',
  {\it Commun.\ math.\ Phys}{\bf  31} (1973) 83-112
    
 
    
\bibitem{Fr78}
 Juerg~Froehlich,
  ``An introduction to some topics in Constructive QFT'',
  Springer-Verlag, New York, (1978)
 
\bibitem{Riv00}
 Vincent~Rivasseau,
  ``Constructive Field Theory and Applications: Perspectives and Open Problems'',
  {\it J.\ Math.\ Phys.} {\bf 41} (2000) 3764-3775

\bibitem{Sim74}
	Barry Simon.
	The P($\phi$)2 Euclidean (Quantum) Field Theory.
	{\it Princeton Unviersity Press}
	(1974)	



\bibitem{Ash91}
	A. Ashtekar.
	Lectures on non perturbative canonical gravity.
	{\it Word Scientifitc}
	(1991)	

\bibitem{Rov04} 
C. Rovelli. {\it Quantum Gravity}, (Cambridge University
Press, Cambridge, 2004).


\bibitem{Thi07} Thiemann. {\it Modern Canonical Quantum General
Relativity}, (Cambridge University Press, Cambridge, 2007).
[gr-qc/0110034]

\bibitem{GS13}
	K. Giesel and H. Sahlmann.
	From Classical To Quantum Gravity: Introduction to Loop Quantum Gravity.
	(2013),
	[arXiv:1203.2733v2]    



\bibitem{KT91} 	
K. V. Kuchar, C. G. Torre. 
Gaussian reference fluid and interpretation of quantum geometrodynamics.
{\it Phys. Rev.} {\bf D43} (1991) 419-441.

\bibitem{BK95}
 J. Brown and K. Kucha\v{r}.
Dust as a standard of space and
time in canonical quantum gravity.
{\it Phys. Rev.} {\bf D51} (1995), 5600-5629. [gr-qc/9409001]

\bibitem{HP11}
	V. Husain and T. Pawlowski.
	Time and a physical Hamiltonian for a quantum gravity.
	{\it Phys. Rev. Lett.} {\bf 108}
	(2011)
	[arXiv:1108.1145v2]


\bibitem{GT12}
	K. Giesel and T Thiemann.
	Scalar Material Reference Systems and Loop Quantum Gravity.
	(2012)
	{\it Class. Quant. Grav.} {bf 32}
	(2015)
	[arXiv:1206.3807v2]
	
	

\bibitem{WK73}
	K.G. Wilson and J. Kogut.
	The renormalization group and the $\epsilon$ expansion.
	{\it Physics Reports}
	{\bf 12}
	(1974)

\bibitem{Wil75}
	Kenneth G. Wilson.
	The renormalization group: Critical phenomena and the Kondo problem.
	{\it Rev. Mod. Phys}
	{\bf 47}
	(1975)
	

\bibitem{ReuSau07}
	M. Reuter, F. Saueressig.
	Functional Renormalization Group Equations, Asymptotic Safety and Quantum Einstein Gravity.
	{\it In  Ocampo, Pariguan, Paycha (Eds.), Geometric and Topological Methods for Quantum Field Theory} Cambridge University Press.
	(2010)
	[arXiv:0708.1317]
\bibitem{Perc10}
	R. Percacci.
	A short introduction to asymptotic safety.
	[arXiv:1110.6389]

\bibitem{ReuSau12}
	M. Reuter. F. Saueressig.
	Quantum Einstein Gravity.
	{\it New Journal of Physics} {\bf 14}
	[arXiv:1202.2274]
	(2012)
	
	
\bibitem{Eich17}
	Astrid Eichhorn. Status of the asymptotic safety paradigm for QG an matter
	[arXiv:1709.0369]	
	
	


\bibitem{BD09}
	B. Bahr, B Dittrich.
	Improved and Perfect Actions in Discrete Gravity.
	{\it Phys Rev D}
	{\bf 90}
	(2009)
	
\bibitem{BDS11}
	B. Bahr, B. Dittrich, S. Steinhaus.
	Perfect discretization of reparametrization invariant path integrals
	{\it Phys. Rev. D}
	{\bf 83}
	(2011)

\bibitem{DMS14}
	B. Dittrich, S. Mizera, S. Steinhaus.
	Decorated tensor network renormalization for lattice gauge theories and spin foam models.
	{\it New J. Phys} {\bf 18}
	(2016)
	
\bibitem{BS17}
	B Bahr, S. Steinhaus.
	Hypercuboidal renormalization in spin foam quantum gravity.
	{\it Phys Rev D}
	{\bf 95}
	(2017)


\end{thebibliography}
\end{document}